# About the domino problem in the hyperbolic plane, a new solution

Maurice Margenstern,
Université Paul Verlaine − Metz,
LITA, EA 3097, IUT de Metz,
Île du Saulcy,
57045 METZ Cédex, FRANCE,
*e-mail*: *margens@univ-metz.fr*

November 20, 2018

**Abstract**

In this paper, we improve the approach of the previous paper about the domino problem in the hyperbolic plane[8, 9]. This time, we prove that the general problem of tiling the hyperbolic plane with *à la* Wang tiles is undecidable.

## 1 Introduction

The question, whether it is possible to tile the plane with copies of a fixed set of tiles was raised by Wang, [15] in the late 50's of the previous century. Wang solved the *partial* problem which consists in fixing an initial finite set of tiles: indeed, fixing one tile is enough to entail the undecidability of the problem. The general case, later called the **general problem** in this paper, without condition, in particular with no fixed initial tile, was proved undecidable by Berger in 1964, [1]. Both Wang's and Berger's proofs deal with the problem in the Euclidean plane. In 1971, Robinson found an alternative, simpler proof of the undecidability of the general problem in the Euclidean plane, see [13]. In this 1971 paper, he raises the question of the general problem for the hyperbolic plane. Seven years later, in 1978, he proved that in the hyperbolic plane, the partial problem is undecidable, see [14]. Up to now, and as far as I know, the general problem remained open.

In this paper, we give a proof that the general problem is also undecidable in the case of the hyperbolic plane. This paper improves a previous paper where the so called "generalized origin-constrained" problem was proved undecidable. In the setting of that paper, the set of origins of computations is infinite but is not dense in the hyperbolic plane, so that it is not the general problem. It is more general than the origin-constrained problem solved by R. Robonson in his 1978 paper, see [14].

Our present proof turns out to have important similarities with the proofs of Berger and Robinson. However, it is more complex for what is the definition of the starting grid which is simply a square grid in the case of the solutions for the Euclidean plane. Our proof combines basic features of both Berger's and Robinson's proof in the grid of the hyperbolic plane which we indicate. For completeness, we also summarise the whole approach of the proof of the undecidability of the general problem.

In this paper, we intensively use the **mantilla**, the grid which we constructed in our first paper, [8, 9]. We refer the reader to these papers, especially to the technical report [9] where the basic properties are proved. This report is available at the following address:

`http://www.lita.sciences.univ-metz.fr/~margens/hyp_dominoes.ps.gzip`

Here, we simply remember the construction and the properties which we use. Then, in the second section, we introduce new properties which will allow us to define the general construction of infinitely many finite domains of increasing sizes in which computations can be organized. The description of the domains is dealt with in section 2. Section 3 gives the last features needed to organize the computations and gives the proof of the main theorem of this paper:

**Theorem 1** *The general problem of tiling the hyperbolic plane is undecidable.*

Here, we do not remind the definitions given for the mantilla nor its properties. We refer the reader for that to the above indicated technical report, as well as to the first version of this paper deposited on arXiv site, see [8]. We shall start with the new ingredients of the proof which are the following: a new property of the mantilla which allows us to define the analog of horizontal lines in a Euclidean plane, a new model for the definition of the computing area which comes from considerations on brackets. Then, we implement the model in the Euclidean plane. The heart of the proof consists in adapting this model to the hyperbolic plane, providing us with a set of tiles which forces its construction in which we have infinitely many areas of infinitely many sizes.

Due to the limited space of this paper, we give only definitions and results. The proofs are given in a technical report to be found at the URL of the author, see [10].

# 2 New ingredients

## 2.1 The levels and the isoclines

In this section, we introduce a new object with respect to [9], the **isoclines**.

In [9], we defined **levels**. Informally, they recursively connect the sons of flowers. By extending the spanning tree of the splitting associated with the flowers we introduced, see [9], it is possible to define a **carpet** structure which is a kind of limit of what we obtain in a tree when the root of the tree is thrown away to infinity. This structure is studied in [7] in the case of the pentagrid and of the heptagrid and we refer the reader to this paper for properties of these tilings and their proofs. As we shall not use these levels, we do not give their formal definition here. Now, the grid admits another notion of **level** induced by the structure of carpet of [7] which comes from the levels of a Fibonacci tree: the set of nodes which are at the same distance from the root, where the distance from the root to a node $\nu$ is defined by the number of nodes on the path from the root to $\nu$.

As noted in [9], these levels are different from those of the mantilla, although they somehow match with them: in particular, applied to a tree of the mantilla, these levels are symmetric as those of the mantilla. But these levels are different: inside a fixed Fibonacci tree, the maximal distance of a level of the mantilla with the Fibonacci level with which it coincides at the borders of the tree is at most linear in the depth of the Fibonacci level in the tree. However, as defined in [9], these Fibonacci levels have a defect: when we go from a tree to another, the levels do not match, even when the new tree is contained in the other one.

In this paper, we give another construction of the Fibonacci carpet which will solve the problem. Consider trees rooted at a black son and call them **black trees**. We can represent an ordinary Fibonacci tree as a stack of black trees. In [9], this decomposition is used to define the **chords** of the harp. Now, if we systematically apply this decomposition, we obtain a subfamily of Fibonacci levels which we call **isoclines** to differentiate them from the general case. Now, the isoclines match from one tree to another: not only when one tree is contained in the other, but also when the areas have no intersection.

The proof of this important property relies on the following statement:

**Lemma 1** *If the $F$-son $\sigma$ of a $G$-flower is a black node, all other $F$-sons of a $G$-flower inside the tree of the mantilla rooted in $\sigma$ are black nodes. Also, the* **8***-centres are always black nodes.*

The proof is obtained by structural induction on the flower structure. It is illustrated by a the pictures of figure 1, below, for a black $F$-centre and a $G$-centre. The other cases can be obtained from these ones. Also see direct illustrations in [10].

In the figure, convex arcs cross the tiles. In black nodes, the convex part is turned to the top: $\frown$; and in the white nodes it is turned to the bottom: $\smile$. The arcs joins mid-points of the concerned edges. In section 3, we shall precisely look at the incidence of figure 1 on the tiles of the mantilla.

An **isocline** is defined as a maximal set of tiles which can be connected by a sequence of such arcs where two consecutive arcs share a common mid-point of an edge.

This proves that the isoclines of different trees match. Indeed, we precisely have:

**Lemma 2** *When an isocline cuts a tree of the mantilla, it cuts its borders at tiles which are equidistant to the root. In particular, the trace of the isoclines inside a tree of the mantilla are the levels of the tree.*

Now, the isocline have also an important property. Indexing the isoclines in a tree by non-negative integers, the root being on the isocline 0, we have:

**Lemma 3** *Let $T$ be a tree of the mantilla with its root on the isocline $0$. There is an $F$-son of a $G$-flower on the isocline $5$ which is within $T$.*

The proof is an easy corollary of the pictures of figure 1. As a corollary of this important property, we have:

**Lemma 4** *Starting from the isocline $4$ of an* **8***-centre $A$ on the isocline $0$, there are $F$-sons of a $G$-flowers on all the isoclines. Moreover, up to the isocline $10$, there are such nodes at a distance at most $20$ from $A$.*

Informally, the isoclines are these horizontals for the whole plane which we need for our construction. The verticals are simply the rays which constitute the border of the trees of the mantilla. This may be surprising as, strictly speaking, the verticals we have just defined are rays, not lines. This is a specific situation of the tiling. There can be such verticals in the hyperbolic plane, for instance the lines issued from a common point at infinity. But in this case, these lines cannot give rise to a continuous set of tiles because of the diameter of a tile. In fact, in the tiling, the trace of such parallel lines are sets of tiles along a ray, and all these rays converge to a common point. This is also the situation of some ultra-threads. Note that there is a realization of the mantilla in which there is a unique line $\ell$ joining **8**-centres such that all the axes of these **8**-centres are supported by $\ell$.

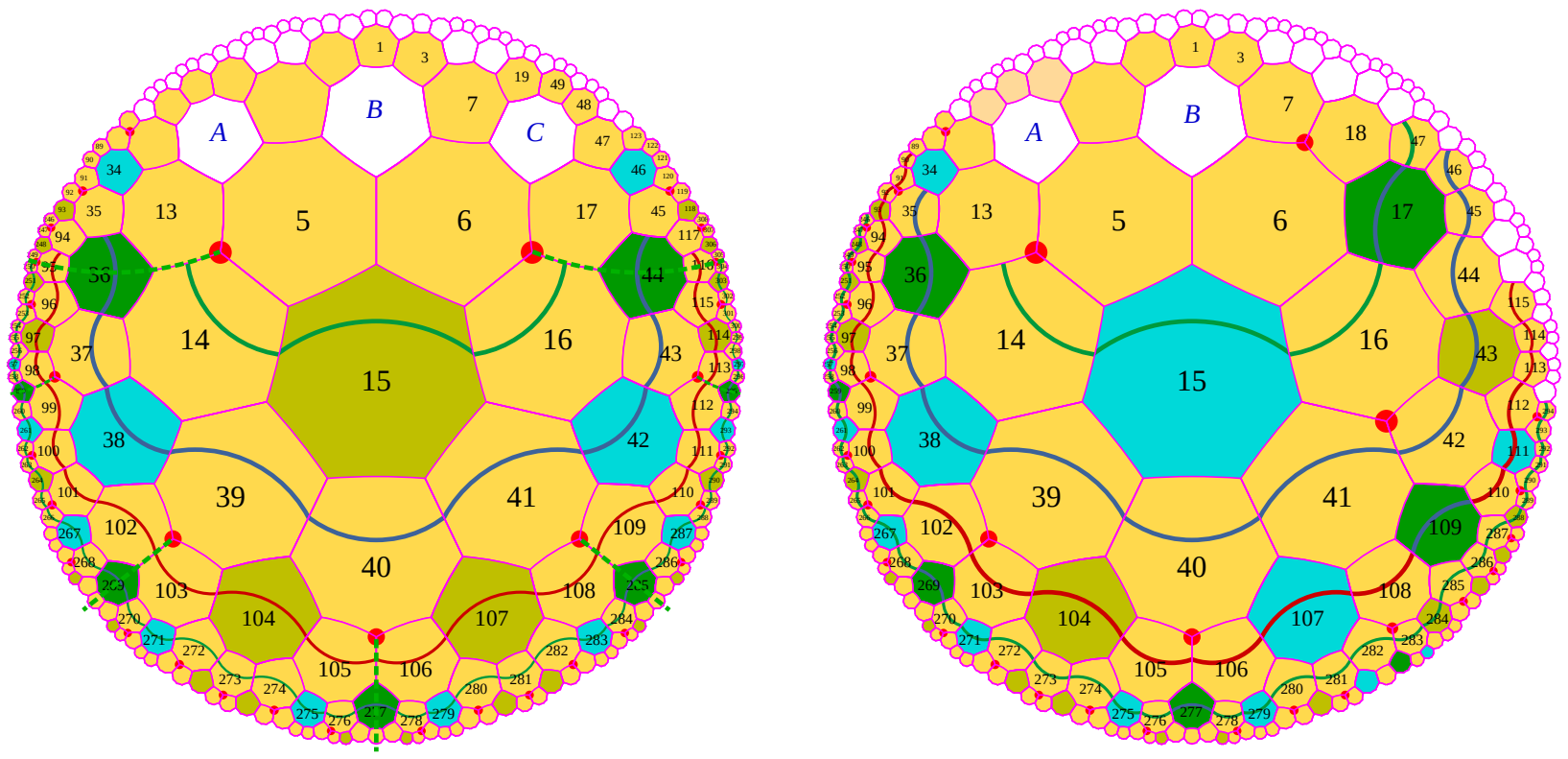

**Figure 1** *The black node property and the levels:*
*On the left-hand side, a black $F$-centre; on the right-hand side, a black $G_\ell$-centre. We can see the case of an* **8***-centre on both figures.*

From now on, we number all the isoclines by indices from 0 to 19, peridocially. We agree that the isocline $n+1$ contains the sons of the nodes of the isocline $n$ for all $n$'s and that the sons of an isocline 19 are on an isocline 0. This defines a direction on the isoclines which goes downwards as the indices increase.

Also from now on, due to lemma 3, we give a special rôle to the isoclines with the numbers 0, 5, 10 and 15 which we shall call the **rows of the mantilla**, simply the **rows**. The definition of tiles to perform this organization raises no difficulty. It is enough to propagate the numbering using the carpet structure. The edge to the father bares the sign **i** while the edges to the son, and only them, bare the sign **i+1**.

Now, we define the **seeds** as the $F$-sons of a $G$-flower which are on a row. When a seed eventually gives rise to a tree, we shall call it **active**. When this is not the case, we call it **silent**. It is important, already now, to notice that not all seeds will be active. However, anticipating on section 3, we indicate that all seeds of the isoclines 0 are active. For the trees emanating from active seeds, all the properties we have seen for the threads and ultra-threads are true for them, provided the set of all trees is restricted to them.

As an important corollary of lemma 4 and the study of the parental tiles for each kind of flower of [9], we have:

**Corollary 1** *For any tile $\tau$ of the mantilla, there is an active seed within a ball of radius* 30 *around $\tau$.*

Now, we turn to the basic process of the construction. This process is also used by Berger and by Robinson in their respective proofs. However, the process, alghough different, is more explicit in Berger's presentation where it is one dimensional, while Robinson directly deals with the two-dimensional extension of the same process. To better understand the process, we look at it from an abstract point of view.

## 2.2 A parenthese on brackets

We shall consider this process from two points of view. In a first sub-section we shall consider the case when the process evolves on the whole line. In a second sub-section, we shall look at what happens if we restrict it to a ray.

### 2.2.1 The bi-infinite case

Consider the following process:

We have a bi-infinite word of the form ${}^{\infty}(RMBM)^{\infty}$. We call this the **row** 0. We define the next **rows** $k$ as follows. Assume that the row $k$ is of the form ${}^{\infty}(R_^{2^k-1}M_^{2^k-1}B_^{2^k-1})^{\infty}$. The word $R_^{2^k-1}M_^{2^k-1}B$ is called the **basic pattern** of generation k or the **$k$-basic pattern**. We superpose these rows one above the other in such a way that the letter $R$ or $B$ of the row $k$+1 is put over the $M$ of a $k$-basic pattern. To better visualize the process, we consider the even rows as written in blue and the odd rows as written in red:

These rows can also be obtained from one another by a cancellation process: on the row $k$, we re-write with blanks the letters $R$ and $B$ and the $M$ letters are re-written $R$ or $B$ according to the following rule. At random, we fix one $M$ at the mid-point of a basic pattern and then, we replace this $M$ by $R$. The next $M$ to the right is left unchanged and the second one is replaced by $B$ and the third one is again left unchanged. Starting from the following $M$ we repeat this process periodically. We also repeat the reverse sequence of actions to the left of the initially chosen $M$.

Now, the letters define positions in a natural way for any subword which we shall consider. We shall call **intervals** the set of positions delimited by the $R$ and the $B$ of a basic pattern. We shall specify blue or red interval according to the colour of the delimiting $R$ and $B$. We shall also consider the set of positions defined by a $B$ followed by the next $R$. We shall call this a **silent interval**, also specifying red or blue, depending on the colour of the delimiting $B$ and $R$. When speaking of silent intervals, the ordinary one will be called **active**. At this point, we notice that active and silent intervals of the same generation have the same length and the same stucture of proper sub-intervals, whether active or silent.

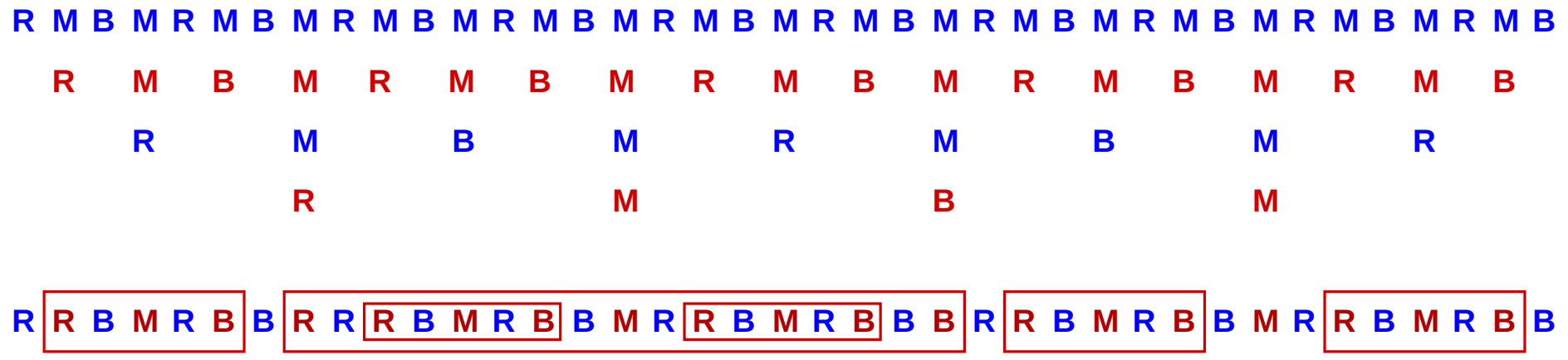

**Figure 2** *The abstract rows and the transparent letters: inside the long red frame the transparent letters which are not in the smaller red frames.*

Now, assume that we superpose the rows of the generations from 0 to $2n+1$. In this superposition, we assume that the blanks and the blue letters are transparent. We are interested in counting the number $f_n$ of letters inside a red active interval of the generation $2n+1$ which are not contained in a red active interval of a previous generation. This is illustrated in figure 2. Such letters will be called **free**. Note that, from this definition, the letters which are the ends of a red interval cannot be free.

We have the following property which can easily be derived from figures 2 and 3 by induction.

**Lemma 5** *The number of free letters inside a red interval of the generation* $2n-1$ *is* $2^n+1$.

Now, we can also ask the same questions for the blue intervals, considering that now, blank and red letters are transparent. We easily get the following result, again by induction:

**Lemma 6** *The number of free letters inside a blue interval of the generation* $2n$ *is* $1$ *for* $n > 0$.

The result of lemmas 5 and 6 explain why we shall consider the red intervals in our construction.

The free letters also have another interesting property in the red intervals.

**Lemma 7** *Let $a_1, \ldots, a_{2^n+1}$ be the positions of the free letters in a red interval of the generation $2n+1$, the first position in such an interval being 1, the ends being not taking into account. Then $a_{i+1}-a_i \geq 2$ except for $i=2^n$ and $i=2^n+1$ for which $a_{i+1}-a_i=1$.*

The proof is done by induction on $n$, the property being already true for the generation 1. Note that in the proof of lemma 5, we had to look at halves of intervals of the form $]BMR[$. In such intervals, the central $M$ is never counted as well as the ends. Accordingly, by induction, the contribution of a half consists of isolated free letters.

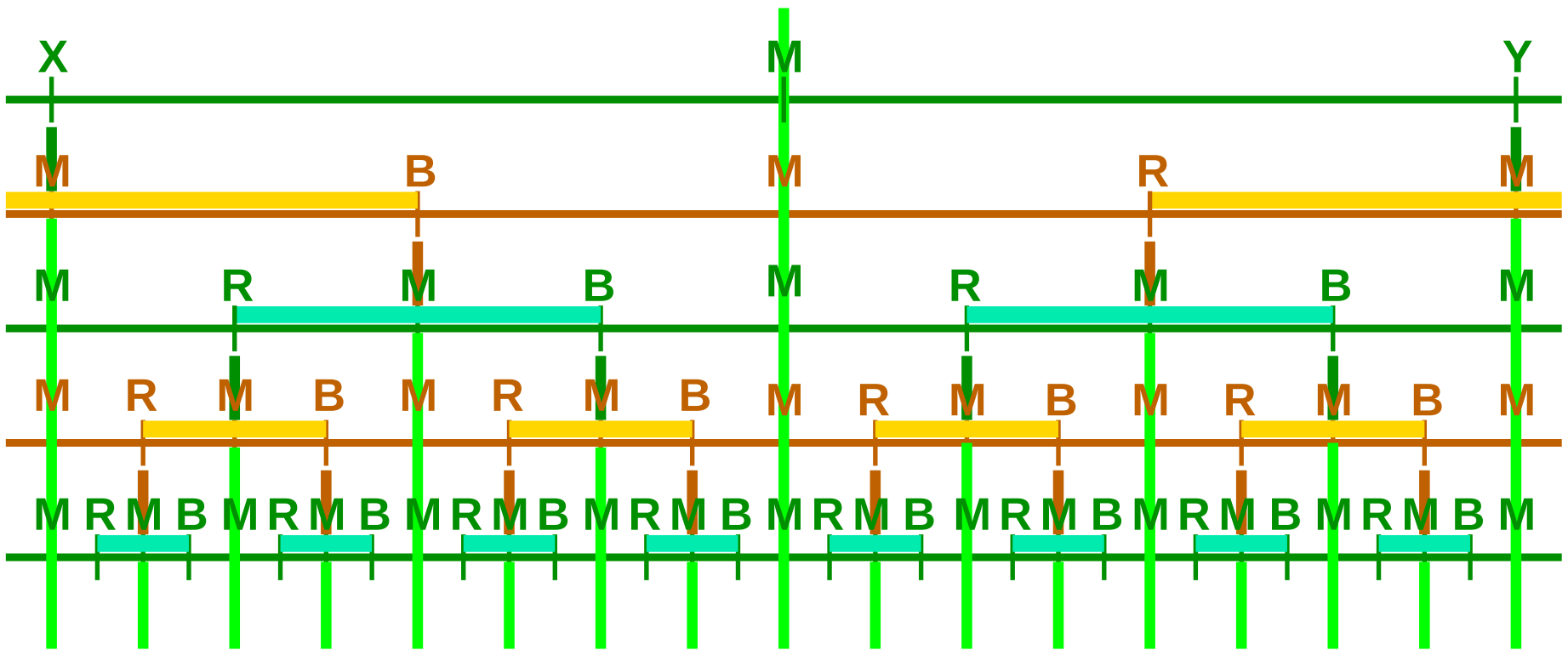

**Figure 3** *The silent and active intervals with respect to mid-point lines. The light green vertical signals send the mid-point of the concerned interval to the next generation. The colours are chosen to be easily replaced by red or blue in an opposite way. The ends $X$ and $Y$ indicate that the figure can be used to study both active and silent intervals.*

We can describe the structure of the red active intervals contained in a bigger red active interval and of the free letters with better precision. First, it is not difficult to see that the free letters come from letters which are already present in the generation 0:

**Lemma 8** *Let $I$ be a red active interval of the generation $2n+1$, with $n > 0$. Let $x$ be a letter $R$, $M$ or $B$ belonging to a blue active interval of the generation $2m$, with $0 < m \leq n$. Then $x$ cannot be a free letter.*

The proof is also done by induction. Now, still by easy inductions, we can describe the positions of the red active intervals and of the free letters contained in a given active interval are given by the following lemma:

**Lemma 9** *Let $I$ be a red active interval of the generation $2n+1$. Then, $I$ can be split into its ends, its free letters and a finite set of interval which exactly contains $2^k$ intervals of the generation $2(n-k)+1$. Similarly, if $J$ is a blue active interval of the generation $2n$, then $J$ can be split into its ends, its unique free letter and a finite set of intervals which exactly contains $2^k$ intervals of the generation $2(n-k)$.*

We conclude this sub-section with an additional important information on the silent interval, whose structure is very different from that of the active ones.

Say that a finite sequence $\{I_k\}_{k\in[0..n]}$ of silent intervals is a **tower** if and only if for each $k \in [0..n]$, $I_k$ belongs to the generation $n$ and if the mid-point of these intervals is the same position in $I_n$ which thus, contains all of the members of the sequence. Note that if $\{I_k\}_{k\in[0..n]}$ is a tower, we have $I_k \subset I_{k+1}$ for all $k$ with $0 \leq k < n$. The common mid-point of the silent intervals of a tower is called the **mid-point** of the tower. The last interval in the tower which contains all of them is called the **area** of the tower.

We have the following lemma:

**Lemma 10** *Let $I$ be a silent interval of generation $n$. The silent intervals which it contains or intersect can be partitioned into finitely many towers. The set of the mid-points of these towers contains the mid-point of $I$. For generations $n \geq 2$, it also contains the both ends of $I$. At last, for the towers which are contained in $I$ and whose mid-point is not that of $I$, their area belong at most to the generation $n-2$ and their mid-points are ends of proper intervals of $I$.*

The proof of this lemma is simply by induction on the generation to which $I$ belongs. It can be illustrated by figure 3.

This gives a simple algorithm to construct the active and silent intervals generation after generation, considering that each generation is put on another line which we call **layer**.

Initial step:
$n := 0$;
the current layer is the layer $n$; the active and silent intervals of this generation are determined.

Induction step:

*i*) The mid-point of the active intervals of the generation $n$ which are determined on the layer $n$ send a red signal to the layer $n+1$: it is the current **end signal** which consists of two kinds of signal. Each second signal is called $R$ and the others are called $B$, the initial $R$ being taken at random.

*ii*) The mid-point of the silent intervals of the generation $n$ send a light green signal to the further layers: it is the current **mid-point signal**.

*iii*) The layer $n+1$ stops the end signals. The $R$-signals define the beginning of an active interval of the generation $n+1$ and the next $B$-signal received on the layer defines the terminal end of the considered active interval. The end signals are absorbed by the layer $n+1$. The complement intervals are the silent intervals of the generation $n+1$. The current mid-point signals which meet the mid-point of a silent interval go on to the next layer. The current mid-point signals which reach the mid-point of an active interval will emit the end signal of the next generation.

*iv*) $n := n+1$;

**Algorithm 1** *The algorithm to construct the active and silent intervals, generation after generation.*

We shall go back to this algorithm in the next sub-section.

We conclude this sub-section with a look at the possible realizations of the abstract brackets. From now on, such a realization will be called an **infinite model** of the abstract brackets, **infinite model** for short. At each generation, we have two choices for defining the position of the active intervals of the next generation. Accordingly, this yields continously many infinite models, even if we take into account that if we fix a position which we call 0, there are countably models which can be obtained from each other by a simple shift with respect to 0. However, the different models do not behave in the same way if, for instance, we look at the towers of silent intervals. In this regard, we have two extremal models. In one of them, 0 is the mid-point of an infinite tower of silent intervals. As this model is symmetric with respect to 0, we call it the **butterfly model**. In another model, 0 is always the mid-point of an active interval of each generation. This model is also symmetric with respect to 0. We call it the **sunset model**.

As we shall later have to deal with the butterfly model, let us briefly indicate the following. Let $a_n$, $b-n$ and $c_n$ denote the addresses of the $R$, the $B$ and the $M$ of an active interval $I$ of the generation $n$. Denote by $d_n$ the address of the $M$ which follows the $B$ defined by $b_n$. If $I_{n_1}$ is the active interval of the generation $n$ for which $a_n$ is the smallest positive number, we have $a_{n_1} = 2^n$, $b_{n_1} = 3.2^n$, $c_{n_1} = 2^{n+1}$ and $d_{n_1} = 2^{n+2}$. Note that $a_{n_1}$ also gives the address of the left-hand side end of the silent interval with the smallest positive address in the sunset model.

First, we have the following 0-1 property for the infinite models:

**Lemma 11** *Consider an infinite model of the abstract brackets. We have the following alternative: either for any position $x$, $x$ belongs to finitely many active intervals or, for any $x$, $x$ belongs to infinitely many active intervals.*

The proof relies on the fact that if $x$ belongs to finitely many active intervals, there is a layer $n$ such that if $x$ belongs to an interval of the generation $m$ with $m \geq n$, then the interval is silent. Let $I$ be the silent interval of the generation $n$ which contains $x$. Then, it is plain that the tower to which $I$ belongs is infinite. Otherwise, there would be an active interval containing $x$ and belonging to a generation $k$ with $k > n$, a contradiction. Now, as an infinite tower is unique when it exists, the property holds for any position.

### 2.2.2 The semi-infinite case

Now, we consider what happens if we cut the result of the previous process, as illustrated by figure 3, at some position.

To simplify things, we take the $M$ of a silent interval of the generation 0. We imagine a vertical line $\delta$ starting from $M$, the layers being horizontal, and we say that $\delta$ cuts all the layers. By definition, we forget what happens on the left-hand side of $\delta$ and we look only at what remains on the right-hand side. Moreover, all active intervals which are cut by $\delta$ are removed. What remains will be called a **semi-infinite model**

There are a lot of realizations of the semi-infinite model. Say that when a position is contained by an active interval, it is covered by this interval.

Consider an infinite model of the abstract brackets, fix a position $x$ and focus on the semi-infinite model defined by the cut at $x$. Then we have:

**Lemma 12** *Consider an infinite model of the abstract brackets and the position $x$ of a cut. In the semi-infinite model defined by the cut at $x$, for any position $y$ after $x$, $y$ is covered by finitely many active intervals.*

The proof is straightforward from the following property:

**Lemma 13** *Consider an infinite model of the abstract brackets and the position $x$ of a cut. Let $y$ a position with $x > y$ and assume that in the model, any position is covered infinitely many often. Then, there is an active interval $I$ which contains both $x$ and $y$.*

We omit the easy proofs of these properties which are indicated in [10].

Deep results on the space of all these realizations are given by an acurate analysis to be found in [5]. The interested reader should have a look at this paper.

Before describing the computing domains, we shall make another intermediate step: from abstract brackets, we shall construct interwoven triangles which we define in the next sub-section.

We construct these triangles in a kind of tiling in the Euclidean plane with square tiles, all of the same size. Then, we shall see how to define tiles for constructing these triangles. The next step will be to implement these tiles in the heptagrid, which will turn out to be not that trivial. Some tuning will be needed due to specific features involved by the hyperbolic plane and by the mantilla.

### 2.3 Interwoven triangles

In the Euclidean plane, we fix a vertical line which we call the **axis**. Then, we define equidistant lines, perpendicular to the axis which we call the **rows**. We number by the elements of $\mathbb{Z}$. By definition, the numbering increases as we go downwards through the rows.

Our goal is to repeat the bracket construction on triangles which will be isoceles. Intervals of the previous sub-section are the projections of the triangles on the axis. There will be blue and red triangles and triangles of the same generation are equal. We want the same connection between triangles as between brackets. We can visualize this as displayed by figure 4.

The layers of algorithm 1 can be considered as parallel lines lying in the same plane $\Pi$. We can imagine that each layer is the intersection of a plane which is orthogonal to $\Pi$. Say that these planes are vertical. Now, we consider that the triangles of a generation are in the vertical plane defined by the layer. In this setting, the intervals on a layer are the projections of the triangles of the plane and the red triangles are projected on the red intervals and the same for the blue triangles with the blue intervals. If we orthogonally project all these planes on the first one, we get the interwoven triangles. Now, note that, here also, we shall distinguish between the triangles whose projection is an active interval and those whose projection is silent. The formers are simply called **triangles** while the latters are called **phantoms**. For properties which are common to both triangles and phantoms, we shall speak of **trilaterals**. By definition, the **status** of a trilateral indicates whether it is a triangle or a phantom.

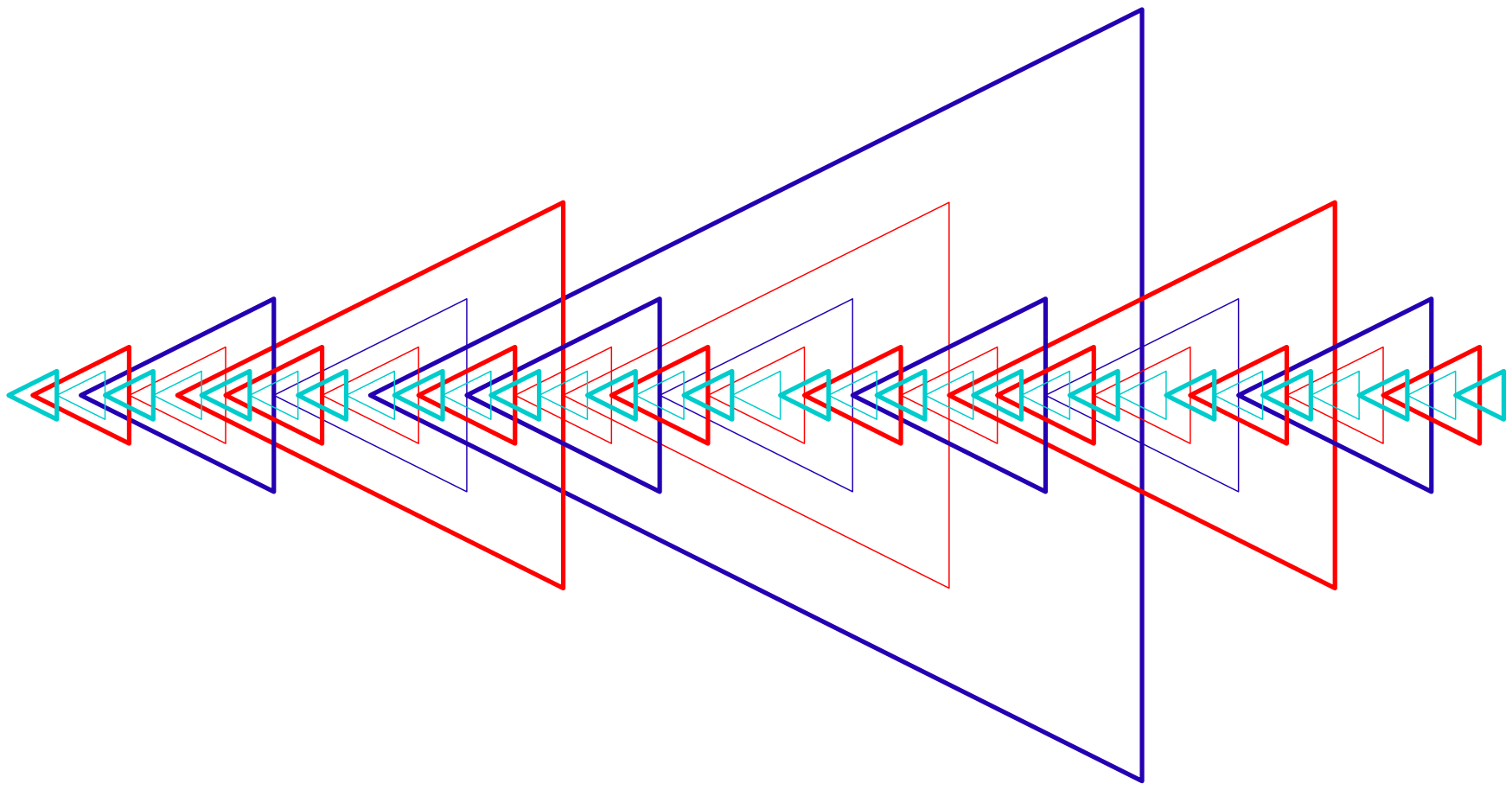

**Figure 4** *An illustration for the interwoven triangles.*

We can now transfer the principles of algorithm 1 to this setting in order to obtain another algorithm, allowing to construct the trilaterals. This algorithm will be at the basis of our construction for the solution of the general problem.

Initial step:

*i*) Put the vertex of a blue triangle on the intersection of each line $4n+1$, $n \in \mathbb{Z}$, with the axis. The basis of the corresponding triangles are on the lines $4n+3$, and the height of each triangle lies on the axis. Each triangle is uniquely determined by $n$. This is the generation 0 of triangles. The phantoms of the generation 0 are determined by the same lines. But the vertex of a phantom is a line $4n+3$ while its basis is on the following line $4n+5 = 4(n+1)+1$.

*ii*) Call **mid-distance line** the line $4n+2$ for a triangle and the line $4n$ for a phantom. Call **mid-point** of an interwoven triangle the intersection of the mid-distance line with the axis. Of course, the mid-point of an interwoven triangle is not its isobarycentre.

Induction step:

*i*) The construction of the trilaterals of the generation $n$ is completed. In particular, we know the mid-distance line of each of them. The trilaterals of an even generation are blue. Those of an odd generation are red. All trilaterals of the same generation can be generated by taking one of them and shifting it along the axis by repeating an appropriate displacement, say $d_n$.

*ii*) Take at random a triangle of the generation $n$ and denote it by 0. The next triangle of the same generation on the axis, downwards, is denoted by 1 and so on. Similarly, negative numbers are used upwards. The mid-distance lines of the trilaterals of the generation grow outside the trilateral, bearing a green signal. They also grow as a basis of a triangle in the triangles of the generation $n$ with an odd index, as a basis of a phantom in the triangles of the generation $n$ with an even index. The mid-point of the triangles with an even index emit legs of the triangles of the generation $n+1$. The mid-point of the triangles with an odd index emit legs of the phantoms of the generation $n+1$. The legs of the trilaterals of the generation $n+1$ is parallel to the legs of the trilaterals of the generation $n$. The legs grow until they meet the nearest mid-distance line, bearing a green signal. Afterwards, they grow until they meet the nearest basis. When the legs meet the green signal, they stop it if they are legs of a triangle and they let it grow outside the trilateral which they define when they are legs of a phantom. In all cases, the basis signal which accompanies a green signal goes on after meeting the legs.

*iii*) $n := n+1$;

**Algorithm 2** *The algorithm to construct the trilaterals.*

Note that the trilaterals are constructed as isoceles triangles, with their heights supported by the axis and with their legs, parallel to the legs of the trilaterals of the generation $n$. We also notice that the distance between two consecutive trilaterals of the generation $n+1$ and of the same kind is now $d_{n+1} = 2d_n$. By definition, the trilaterals of the generation $n+1$ have a colour which is opposite to the colour of the trilaterals of the generation $n$. By definition, red and blue are opposite of each other. Also note that the mid-distance signals of the generation $n$ whose green part is stopped by a triangle of the generation $n+1$ grow outside this triangle: either as a basis of a triangle or a phantom of the generation $n+2$.

Now, we note the following property:

**Lemma 14** *The triangles of the generation* $n{+}2$ *do not meet the triangles of the generation* $n$*. Accordingly the area of a triangle of the generation* $n$ *is contained in the area of some triangle of the generation* $n{+}2$*.*

Note that the property does not hold for the phantoms. For the phantoms, we can collect them into towers, exactly as this was done for the silent intervals. We have:

**Lemma 15** *The phantoms of a tower have the same mid-point. This mid-point may be the vertex of a triangle or the mid-point of a basis of a triangle. In a tower, the colours of the phantoms it contains alternate.*

Note that the green-signal of the mid-distance line of a phantom goes on outside the phantom as long as it meets legs of phantoms. It can be stopped only by the legs of a triangle. There can be realizations of the trilaterals where such a line never meet the legs of a triangle. This corresponds to the butterfly model in terms of the intervals obtained by projection of the triangles on the axis. If this is the case, this happens for one mid-distance line only, and so there is a unique realization with this property, up to the colour of the basis which accompanies the green signal.

It does not follow from lemma 14 that the triangles cannot meet. In fact triangles of different colours may meet. From the very construction given by algorithm 2, a given triangle $T$ meets at least two triangles of the opposite colour and which belong to the previous generation: one around its vertex, the other crossing its basis. For this second triangle, its mid-point is the mid-point of the basis of $T$.

Now, the same triangle may meet another one: if its mid-point is the vertex of a triangle $T_1$ of the next generation, $T_1$ will meet the basis of $T$. Now, if its mid-point is the vertex of a phantom $P$, then the mid-distance line of $T$ also supports the basis of a triangle $T_1$ of the next generation and this basis meets the legs of $T$. And so, a triangle always meets three triangles: two of the previous generation and one of the next one. Note that from the definitions, all the triangles met by $T$ are of the opposite colour.

And so, when the basis of a **triangle** cuts the legs of another **triangle**, the intersection occurs at the mid-point $M$ of the legs or at the mid-point between the vertex and $M$.

Now, the situation with the phantoms is different as we already noticed. The abstract bracket may help us in this regard. Note that the order from the left to the right coresponds to an inclusion order for the angles associated to the trilaterals attached to the intervals. Namely, if $x$ and $y$ respectively are two left-hand side ends of two intervals $I$ and $J$, $x$ and $y$ are associated to vertices $X$ and $Y$ of trilaterals $F_X$ and $F_Y$ and the angle at $X$ contains the angle at $Y$. Of course, this does not mean an inclusion between $F_X$ and $F_Y$. This depends on the relative positions of the other ends of $I$ and $J$, say $u$ and $v$ respectively. The connections between the trilaterals are summarized by table 1, below, also see figure 4.

From the tower structure and the position of the mid-points of the towers and the above remark, we can conclude that the phantoms of a tower around the vertex $V$ of a triangle meet the triangle: the bases of these phantoms will cut the legs of the triangle. But the legs of these phantoms will also meet the basis of the phantom whose mid-point is $V$. If the mid-point of the tower is the mid-point $M$ of the basis of a triangle, we have a dual situation: the legs of the phantoms will cut the basis of the triangle but also the bases of these phantoms will cut the phantom with vertex $M$. For the towers inside a given triangle or phantom, we have the same relations with triangles and phantoms of an older generation due to the position of the mid-point of a tower.

Note that contrary to what happens with triangles, when the basis of a phantom cuts the legs of a triangle, this does not occur at the two points which we noticed. As it can easily be inferred from table 1 and from the proof of lemma 5, see also figure 3, this occurs inside the **first half** of the legs, *i.e.* the part which goes from the vertex of the triangle to the mid-point of the leg. More precisely, it occurs between the vertex and the mid-point of this first half. The same property is also true for a phantom, as the greatest member of a tower might be replaced by a triangle. We may summarize this study by lemmas 16 and 17, below.

| | |
|---|---|
| $u < y$ | $ar(F_X) \cap ar(F_Y) = \emptyset$ |
| $y < u < v$ | the basis of $F_X$ cuts the legs of $F_Y$ |
| $v < u$ | $ar(F_Y) \subset ar(F_X)$ |

**Table 1** *The connections between the trilaterals according to the relations between their projections. Remember that $ar(F)$ is the area of the trilateral $F$.*

**Lemma 16** *Between trilaterals, the crossings occur between legs for one figure and the basis for the other figure.*

**Lemma 17** *A leg of a trilateral is never cut inside the open interval delimited by the mid-distance line and the basis of the trilateral.*

The goal of the next sub-section is to prove the following result:

**Lemma 18** *The interwoven triangles can be obtained by a tiling of the plane which is finitely generated and submittted to the constraint that rotations and symmetries of the prototiles are ruled out.*

## 2.4 Tiles for the interwoven triangles

### 2.4.1 The background

Our tiles are square tiles, but they are not placed by tessellation. We may imagine them as put in isometric rows and then, we shift all odd rows by half the side of the square.

It is not difficult to imagine tiles forcing this property. An easy solution is given in [10]. Here, we consider this point as granted.

In the rows defined by this tiling, each second one will be passive: it will simply transmit signals unchanged in order to define simply the slope of the legs of the triangles. This is also not difficult to implement, see [10].

### 2.4.2 Implementation of algorithm 2

Now, we have to translate the construction ruled by algorithm 2 into the tiles we have just introduced in this sub-section. As we have a rather important number of tiles, especially for the implementation in the hyperbolic plane, we shall constitute the tiles we need from **elementary patterns** which are represented by Figures 5, 6, 7 and 8. Then, we shall define a few **operations** on the elementary patterns from which we shall obtain the tiles which we propose for the implementation of algorithm 2.

We remark that, for the legs, we distinguish between the two halves by a difference of **hue** inside a colour. For blue-0 and simple blue trilaterals, the first half has a **dark** colour and the second half has a **light** one. For red trilaterals, the convention is opposite: the first half has a **light** colour and the second half has a **dark** one. Next, the difference between triangle and phantom is marked by the width of the stroke: legs of a triangle are **thick**, those of a phantom are **thin**.

We shall denote these patterns by **names** as follows, first for the legs and then for the bases:

| | | | | | | | |
|---|---|---|---|---|---|---|---|
| $Lb_0tul$ | $Lb_0tur$ | $Lb_0tbl$ | $Lb_0tbr$ | $Lb_0\varphi ul$ | $Lb_0\varphi ur$ | $Lb_0\varphi bl$ | $Lb_0\varphi br$ |
| $Lb_ntul$ | $Lb_ntur$ | $Lb_ntll$ | $Lb_ntlr$ | $Lb_n\varphi ul$ | $Lb_n\varphi ur$ | $Lb_n\varphi ll$ | $Lb_n\varphi lr$ |
| $Lrtul$ | $Lrtur$ | $Lrtll$ | $Lrtlr$ | $Lr\varphi ul$ | $Lr\varphi ur$ | $Lr\varphi ll$ | $Lr\varphi lr$ |
| | $Bb_nt$ | $Bb_n\varphi$ | $Bb_0t$ | $Bb_0\varphi$ | $Brt$ | $Br\varphi$ | |

*a.* leg patterns:

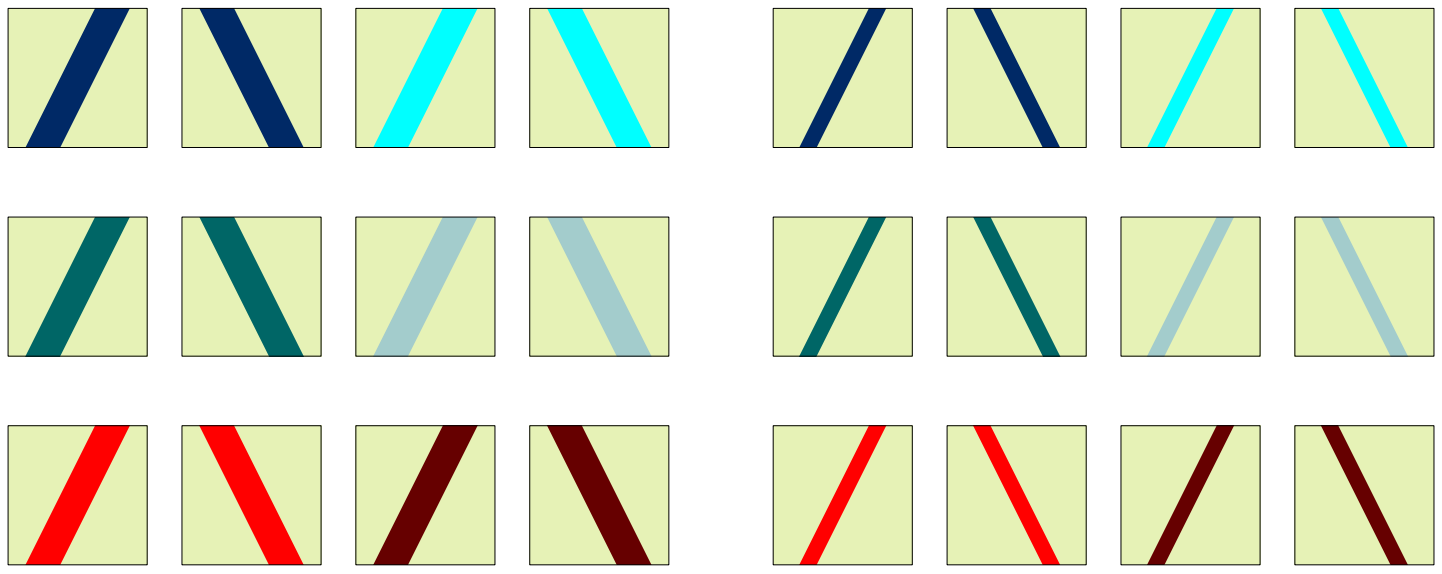

**Figure 5** *The elementary patterns to define the legs of triangles and phantoms according to their colour and also taking into account the important difference between the first and the second half of legs.*

*b.* basis patterns:

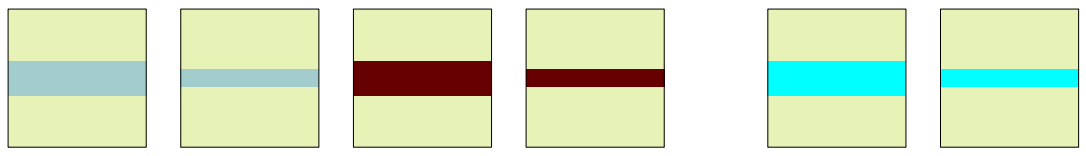

**Figure 6** *The elementary patterns for the bases.*

For the join-patterns which already combine elements of the elementary patterns, we define the following names, first for vertices, then for mid-points and then for corners:

| | | | | | |
|---|---|---|---|---|---|
| $Vb_0t$ | $Vb_0\varphi$ | $Vb_nt$ | $Vb_n\varphi$ | $Vrt$ | $Vr\varphi$ |
| $Mb_0tl$ | $Mb_0tr$ | $Mb_ntl$ | $Mb_ntr$ | $Mrtl$ | $Mrtr$ |
| $Mb_0\varphi l$ | $Mb_0\varphi r$ | $Mb_n\varphi l$ | $Mb_n\varphi r$ | $Mr\varphi l$ | $Mr\varphi r$ |
| $Cb_0tl$ | $Cb_0tr$ | $Cb_ntl$ | $Cb_ntr$ | $Crtl$ | $Crtr$ |
| $Cb_0\varphi l$ | $Cb_0\varphi r$ | $Cb_n\varphi l$ | $Cb_n\varphi r$ | $Cr\varphi l$ | $Cr\varphi r$ |

For the information signals we have the following names:

$Hrl$ $\quad Hrr$ $\quad Hg$ $\quad Hy$ $\quad Z$ $\quad p$

The information signals correspond to what was indicated in the algorithm. The yellow signal marks the free rows inside a red triangle. The green signal indicates the mid-distance row of a trilateral. Remember that inside a triangle, the green signal is stopped by the legs of the triangle. This is not the case for a phantom: the green signal crosses its legs. We have also the horizontal red signals: the right-hand and left-hand side ones which are used inside and outside the red triangles to mark the non free rows.

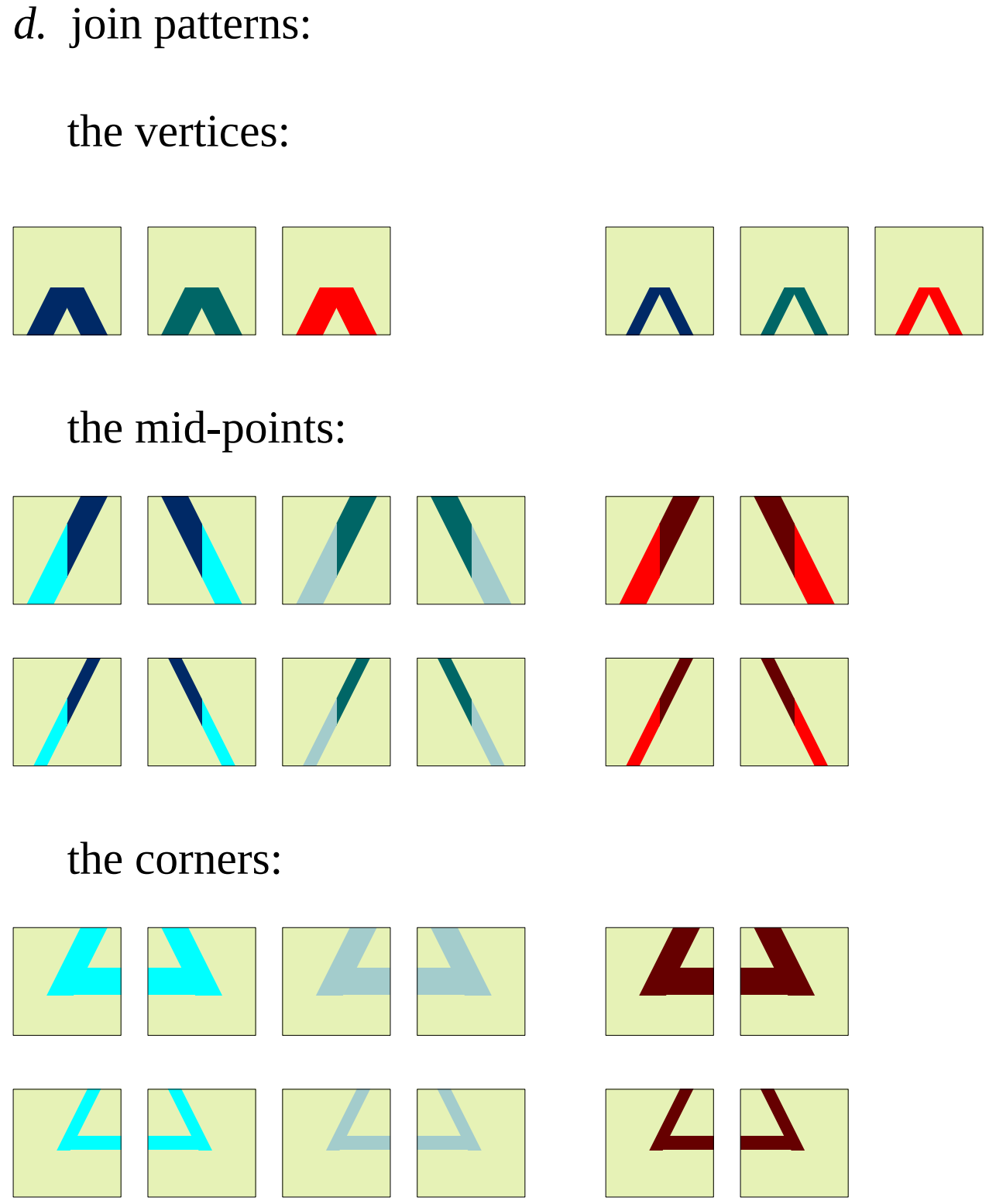

**Figure 7** *The elementary patterns for the vertices, for the mid-point of the legs and for the corners.*

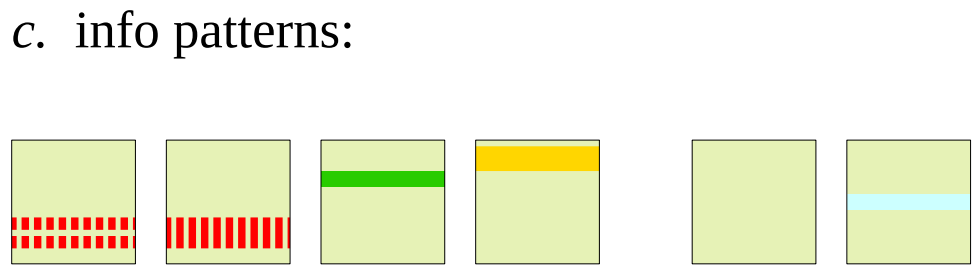

**Figure 8** *The elementary patterns for the horizontal signals used in the construction.*

The patterns of this group can be obtained from the previous ones by means of **operators** which we now define. We introduce four such operators called **masking** operators, denoted by $\mu_L$, $\mu_R$, $\mu_B$ and $\mu_T$, the subscripts meaning *left*, *right*, *bottom* and *top* respectively. The result of $\mu_X(T)$ is to show the part of the tile $T$ defined by $X$. In this way, $L$ defines the left-hand half of $T$, $R$, its right-hand half, $B$ its lower half and $T$ its upper one. All tiles can be obtained by composition of these operators to the previous signals and an operation of **superposition** of the results. We consider the patterns as opaque under the superposition. In principle, this makes this operation non-commutative. However, in the situations we consider in this report, the difference between $A + B$ and $B + A$ is not meaningful. Accordingly, we shall consider the superposition as commutative and, with this convention, it is also associative.

As an example of such a possibility, it is not difficule to see that we can write $V\gamma\tau = \mu_B(\mu_L(L\gamma\tau ul)+\mu_R(L\gamma\tau ur))$, where $A + B$ denotes the superposition of the two tiles $A$ and $B$. In some sense, we can see the names $V\gamma\tau$ as useful shortcuts for much longer expressions.

We can see that the tiles consist in various ways to assemble vertical and horizontal signals. Most often, the horizontal signals involve a basis. The formulas which we shall soon define indicate which signals may accompany a basis.

We shall represent the conditions in which the elementary patterns are combined in order to define tiles by formulas indicating superpositions. Let $A$ and $B$ denote tiles possibly obtained by superposition. We define $\epsilon_a.A + \epsilon_b.B$ as a superposition whose result may be nothing, $A$, $B$ or $A+B$ depending on whether $\epsilon_a$ or $\epsilon_b$ are 1 or 0. The tile $A$, respectively $B$, is present in the superposition if and only if $\epsilon_a = 1$, respectively $\epsilon_b = 1$. We shall often write $\epsilon_\alpha$, for $\alpha \in \{a, b\}$ as $\mathbf{1}_{Cond}$ where $Cond$ is a formula expressing a condition. Note that such a representation can also be used to count how many tiles this poduces. Let $|A|$, $|B$ be the number of prototiles defined by $A$, $B$ respectively. Let $n_a$, $n_b$ and $n_{a\cap b}$ be the number of cases when $\epsilon_a = 1$, $\epsilon_b = 1$ and $\epsilon_a + \epsilon_b = 2$, respectively. Then, we have that $\epsilon_a.A + \epsilon_b.B$ produces $(n_a - n_{a\cap b})\times|A| + (n_b - n_{a\cap b})|B| + n_{a\cap b}\times(|A|\times|B|)$ prototiles. Of course, when the conditions are independent, *i.e.* $n_{a\cap b} = 0$, $\epsilon_a.A + \epsilon_b.B$ produces $n_a\times|A| + n_b\times|B|$ prototiles. As we recursively assume that $A$ and $B$ may be represented in the same way, we can use this formalism to represent the formulas we need to define the tiles and, also, we may use it to count the number of represented tiles. This counting involves a tree-like representation of the formula. If the conditions are independent, the tree is complete. Otherwise, the possible dependencies between the conditions lead to the cancellation of sub-trees in the tree of the formula.

We have not the room to display the tiles, as there is a lot of them, and we shall simply define them by formulas which we collect into tables 9

and 10.

| case of | formula |
|---|---|
| bases<br>$(B)$ 14 | $B\gamma\tau + \mathbf{1}_{\{\gamma=b_0\}}.\mathbf{1}_{\{\tau=\varphi\}}.\epsilon.Hy$<br>$+ \mathbf{1}_{\{\gamma=b_n\}}.(\epsilon_1.(Hg+Hy)+\overline{\epsilon_1}.(\epsilon_2.Hrl+\overline{\epsilon_2}.Hrr))$<br>$+ \mathbf{1}_{\{\gamma=r\}}.(\epsilon_1.Hg+\mathbf{1}_{\{\tau=\varphi\}}.(\epsilon_2.Hrl+\overline{\epsilon_2}.Hrr)))$ |
| vertices<br>$(V)$ 8 | $V\gamma\tau + B\gamma\overline{\tau} + \mathbf{1}_{\{\gamma=b_0\}}.\epsilon.Hy + \mathbf{1}_{\{\gamma=b_n\}}.(Hg + Hy)$<br>$+ \mathbf{1}_{\{\gamma=r\}}.\Big(Hg+Hy+\mathbf{1}_{\{\tau=\varphi\}}.(\mu_L(Hrl)$<br>$+ \mu_R(Hrr))\Big)$ |
| mid-point tiles<br>$(M)$ 36 | $M\gamma\tau\xi+\mathbf{1}_{\{\tau=t\}}.\Big(B\overline{\gamma}\tau_1+\mathbf{1}_{\{\gamma\neq r\}}.(Hg+\mathbf{1}_{\{\tau_1=\varphi\}}.Hr\xi)$<br>$+ \mathbf{1}_{\{\gamma=r\}}.(\mu_\xi(Hr\xi)+\mu_{\overline{\xi}}(Hy))\Big)$<br>$+ \mathbf{1}_{\{\tau=\varphi\}}.\Big(B\gamma_1\tau_1 + \mathbf{1}_{\{\gamma_1\neq r\}}.(Hg + Hy)$<br>$+ \mathbf{1}_{\{\gamma_1=r\}}.(Hg+\mathbf{1}_{\{\tau_1=\varphi\}}.Hr\xi))$ |
| corners<br>$(C)$ 20 | $C\gamma\tau\xi + \mathbf{1}_{\{\gamma=b_0\}}.\epsilon_1.(\epsilon_2.Hy + \overline{\epsilon_2}.Hr\xi)$<br>$+ \mathbf{1}_{\{\gamma=b_n\}}.Hr\xi + \mathbf{1}_{\{\gamma=r\}}.(\mathbf{1}_{\{\tau=\varphi\}}.Hr\xi +$<br>$\mathbf{1}_{\{\tau=t\}}.\mu_\xi(Hr\xi))$ |
| legs of a triangle<br>first part<br>$(L_ut)$ 28 | $L\gamma tu\xi + \Big(\mathbf{1}_{\{\gamma=b_n\}}.(\epsilon_1.(\alpha.Hy + \beta.Hr\xi)$<br>$+ \overline{\epsilon_1}.(B\gamma_1\tau_1$<br>$+ \mathbf{1}_{\{\gamma_1=b_0\}}.Hy$<br>$+ \mathbf{1}_{\{\gamma_1=b_n\}}.\mathbf{1}_{\{\tau_1=\varphi\}}.Hr\xi$<br>$+ \mathbf{1}_{\{\gamma_1=r\}}.\mathbf{1}_{\{\tau_1=\varphi\}}.Hr\xi))$<br>$+ \mathbf{1}_{\{\gamma=r\}}.(\epsilon_1.(\epsilon_2.(\mu_\xi(Hr\xi) + \mu_{\overline{\xi}}(Hy))$<br>$+ \overline{\epsilon_2}.Hr\xi)$<br>$+ \overline{\epsilon_1}.(B\gamma_1\tau_1$<br>$+ \mathbf{1}_{\{\gamma_1=b_0\}}.(\mu_\xi(Hr\xi)+\mu_{\overline{\xi}}(Hy))$<br>$+ \mathbf{1}_{\{\gamma_1=b_n\}}.\epsilon_2.Hr\xi$<br>$+ \mathbf{1}_{\{\gamma_1=r\}}.\mathbf{1}_{\{\tau_1=\varphi\}}.Hr\xi))\Big)$ |

**Figure 9** *Table of the formulas. Note the number of tiles given with the name of the formula.*

In our study of the interwoven triangles, we remember that we mainly had **horizontal** signals on one hand and **vertical** signals on the other hand. Accordingly, our tiles will have the same property. They will consist in signals, vertical and horizontal ones. The vertical signals are constituents of legs of a trilateral, from the vertices to the corners, as can be seen in figures 5 and 7.

| case of | formula |
|---|---|
| legs of a phantom first part<br>$(L_u\varphi)$ 32 | $L\gamma\varphi u\xi + \Big(\mathbf{1}_{\{\gamma=b_n\}}.(\epsilon_1.(\alpha.Hy + \beta.Hr\xi)$<br>$+ \overline{\epsilon_1}.(B\gamma_1\tau_1$<br>$+ \mathbf{1}_{\{\gamma_1=b_0\}}.\mathbf{1}_{\{\tau_1=\varphi\}}.Hy$<br>$+ \mathbf{1}_{\{\gamma_1=b_n\}}.\mathbf{1}_{\{\tau_1=\varphi\}}.Hr\xi)$<br>$+ \mathbf{1}_{\{\gamma_1=r\}}.\mathbf{1}_{\{\tau_1=\varphi\}}.Hr\xi))$<br>$+ \mathbf{1}_{\{\gamma=r\}}.(\epsilon_1.\epsilon_2.Hr\xi$<br>$+ \overline{\epsilon_1}.(B\gamma_1\tau_1$<br>$+ \mathbf{1}_{\{\gamma_1=b_0\}}.\mathbf{1}_{\{\tau_1=t\}}.\epsilon_2.Hy$<br>$+ \mathbf{1}_{\{\gamma_1=b_n\}}.Hr\xi$<br>$+ \mathbf{1}_{\{\gamma_1=r\}}.\mathbf{1}_{\{\tau_1=\varphi\}}.Hr\xi))\Big)$ |
| legs of a trilateral second part<br>$(L_\ell)$ 22 | $L\gamma\tau l\xi + \Big(\mathbf{1}_{\{\gamma=b_n\}\vee\{\tau=\varphi\}}.\epsilon_1.(\epsilon_2.Hy + \overline{\epsilon_2}.Hr\xi)$<br>$+ \mathbf{1}_{\{\gamma=r\}\wedge\{\tau=t\}}.(\epsilon_1.Hr\xi + \overline{\epsilon_1}.(\mu_\xi(Hr\xi)$<br>$+ \mu_{\overline{\xi}}(Hy)))\Big)$ |
| legs :<br>passive tiles<br>$(L_p)$ 24 | $L\gamma\tau\xi + p.$ |

**Figure 10** *Table of the formulas, continued. Note the number of tiles given with the name of the formula.*

In these formulas, $\gamma$ defines the colour of the signal, $\tau$ is the status of the trilateral to which the signal belongs, $\xi$ is the laterality of the signal, if appropriate, and $\pi$ is the position. When we have a two-valued parameter $\omega$, then $\overline{\omega}$ denotes the value which is not taken by $\omega$. In the case of $\gamma$, we define $\overline{\gamma}$ to be $r$ if $\gamma = b_0$ or $\gamma = b_n$ and we define $\overline{\gamma}$ to be $b$ if $\gamma = r$. Conformally to the above definitions of the superposition of formulas controlled by conditions, we interpret $A + \alpha.B$ as $u.A + \alpha.B$ where $u$ is always 1.

At this point, we may sum up all the countings we have performed. We have 184 tiles, as it can easily be found. To these tiles we have to append all the tiles of figure 8. Accordingly, we find 190 tiles.

When we shall turn to the hyperbolic situation it will be absolutely impossible to represent all the prototiles as there are more than 80 times this number, simply to define the impementation of the interwoven triangles.

However, below, figure 11 illustrates the construction of the Euclidean tiling for the first three generations.

We refer the reader to [10] for the proof of lemma 18, as it is simply a thorough checking that the tiles defined by formulas $(B)$, $(V)$, $(M)$, $(C)$, $(L_u\varphi)$, $(L_ut)$, $(L_\ell)$ and the tiles of figure 8 construct the tiling and nothing else. Of course, in this proof, we take into account that the vertices are on the axis and that we must start with a vertex.

We just have a remark about the green signal. This signal grows until it meets a triangle, whatever the colour, red or simple blue. It may happen that the signal will never meet such a triangle. In this latter case, from formula $(M)$, we can see that the green signal is always accompanied by a basis: either of a triangle or of a phantom, either blue or red. Accordingly, this basis induces a unique vertex on the axis for the corresponding row and a trilateral $\mathcal{T}$ will grow starting from there.

It is not difficult to see that $\mathcal{T}$ grows indefinitely: the first part of its legs will never meet a green signal. Indeed, from the correspondence we studied between vertices of trilaterals and ends of the intervals, we know that if a green line is raised inside $\mathcal{T}$, it will be met by a trilateral which is raised by a vertex placed between the vertex of $\mathcal{T}$ and the green line. This can be checked by the study of the butterlfly model, where 0 would be the absissa of the vertex of $\mathcal{T}$ on the axis.

As noted, the model which we implement may be not the butterfly model. When this is the case, the green signal will always eventually meet the leg of a triangle. There may be a contradiction between the tile of formula $(M)$ which has to be put there and the tile of the basis which is involved by the continuation of the green signal in that sense that the colour of the accompanying signal may be different as well as the width of the signal. To solve this contradiction, we fix a rule: the decision of the choice belongs to the leg of the triangle. This means that we may have to change the tiles on this mid-distance line as well as those which depend on the choice performed at the axis. At the considered step of the construction this involves only finitely many tiles and once this is performed, the new tiles will never be changed again. This is why the construction is correct. We simply note that inside a triangle, everything inside the triangle is fixed when the basis of the triangle is constructed as well as the trilaterals of the next generation which it determines. With a phantom, the construction is temporary: it may be changed but once, because at that moment, the part which is changed falls inside a triangle and so, it cannot again be changed.

# 3 The proof of the main theorem

In this section, we first implement the isoclines in the mantilla. Then, we turn to the implementation of the tiles for the interwoven triangles in the hyperbolic plane. There, we shall see the new features entailed by the hyperbolic plane and by the mantilla. As a consequence, a new signal will be introduced as well as a few new tiles. Then, in a third subsection, we shall deal with the definition of the computing areas.

## 3.1 Implementation of the isoclines in the mantilla

From [9], we know that the mantilla can be constructed from a set of 21 prototiles. At first glance, as we have two kinds of nodes to define the isoclines, this would require 42 tiles. This is not the case as, for instance in the **8**-flower, there are tiles which are always black nodes. A careful analysis of the configurations displayed in figure 1 shows that other tiles share this property and that also some of them are always white nodes.

This analysis is reported in table 2 where for each tile, we indicate the edges whose mid-points are ends of the arc of the isocline which crosses the tile. The edges are indicated according to their labels in the tile for centres. For petals, when an edge has no label, we indicate its position by two neighbouring edges or, by an edge and the red vertex, which means that the considered mid-point is on an edge whose one end point is the red-vertex.

Again for the petals, we also give local numbers to the edges. By definition, the red vertex is between 1 and 7 and the numbers are increasing while clockwise turning around the tile. We refer the reader to [10] for a figure which displays the 34 tiles of the mantilla, equipped with the isoclines. Table 2 indicates the exact location of the level signal with respect to the edges of tile in terms of intrinsic properties of the tiles.

edge to the father is the right-hand side edge. Now, give the number 1 to the edge to the father and number the other edges by consecutive numbers from 2 to 7 while counter-clockwise turing around the tile. The isocline signal goes from the edge 3 to the edge 7 in the case of a black node, and it goes from the edge 2 to the edge 7 in the case of a white node. Call such coordinates of an edge **paternal**.

We notice that 5 tiles are always black: **8**, $\overline{5}\circ 77$, $11\circ\overline{3}$, $137\circ$ and $157\circ$. Also, 5 tiles are always white: $1\overline{4}7\circ$, $11\circ 4$, $\overline{66}7\circ$, $1\overline{22}\circ$ and $47\circ 7$. Next, 10 tiles have a black version and a white one: $G_\ell$, $G_r$, $2\circ 77$, $1\circ 1\overline{3}$, $\overline{5}7\circ 7$, $11\circ 6$, $11\circ 2$, $3\circ 77$, $1\circ 15$ and $6\circ 77$. At last, the tile $F$ has three versions: a black one and two white ones. Note that we also indicate the petals of a $G$-centre which are also parental petals of an $F$-centre. As not all black $F$-centre are in this situation, we have to mark those which are. Note that in each case, as a $G$-centre exists in two versions, black and white ones, this induces for each one three petals as one of them has a single version and the other has two ones.

Namely, the petals are $47\circ7$ and $1\circ15$ for a $G_\ell$-flower with $47\circ7$ always white. They are $37\circ7$ and $1\circ14$ for a $G_r$-flower, with $1\circ14$ always white. Now, there is an important difference with [9] where the tiling was also divised so that realizations of the mantilla with an ultra-thread were ruled out. Here we do not exclude such realizations and, more other we have to identify all possible trees of the mantilla. This is why the just mentioned petals are present in the set of tiles with the mark only.

On another hand, we impose that the seeds are taken among the roots of a tree of the mantilla which belongs to an isocline 0, 5, 10 or 15. We shall distinguish between **active** seeds from which a computation will start and **silent** seeds which do not trigger any computation. In sub-section 2.1, we have indicated that all the roots which are on the isoclines 0 are active seeds. We shall later see the selection process to define the the seeds of isoclines 5, 10 or 15 which become active. The implementation of the numbering from 0 to 19 on the set of tiles of table 2 is not difficult. See [10] for details on this subject. We notice an important consequence of the isoclines and of the numbering.

By themselves, togother with the structure of the mantilla, the isoclines prevent to turn a tile: this would entail contradictions between the arcs of the isoclines. Note that we already know from [9] that the structure of the mantilla itself rule out such a possiblity. Now, the numbering, even if it is only periodic, allows to define the directions up and down. Also, the isoclines themselves allow to define what are the directions left and right. However, this does not exactly correspond to what we define by the same words in the Euclidean plane.

There is a simpler way to locate the arcs at the price of a global property. We know that the isoclines induce the structure of a Fibonacci carpet. This means that any tile has a father. In a white node, the edge to the father is the single edge delimited by the ends of the edges joined by the isocline. In a black node, the ends of the edges joined by the isocline define two consecutive edges. If we see these edges as the upper part of the tile, the edge to the father is the right-hand side edge. Now, give the number 1 to the edge to the father and number the other edges by consecutive numbers from 2 to 7 while counter-clockwise turing around the tile. The isocline signal goes from the edge 3 to the edge 7 in the case of a black node, and it goes from the edge 2 to the edge 7 in the case of a white node. Call such coordinates of an edge **paternal**.

| tile | *st* | edges | relative |
|---|---|---|---|
| $F$ | $B$ | 6-2 | 6-2 |
| | $W$ | 6-1 | 6-1 |
| | $W$ | 7-2 | 7-2 |
| **8** | $B$ | $\overline{6}$-$\overline{2}$ | $\overline{6}$-$\overline{2}$ |
| $G_\ell$ | $B$ | 2-$\overline{6}$ | 2-6 |
| $G_\ell$ | $W$ | 1-$\overline{6}$ | 1-$\overline{6}$ |
| $G_r$ | $B$ | $\overline{2}$-6 | $\overline{2}$-6 |
| $G_r$ | $W$ | $\overline{2}$-7 | $\overline{2}$-7 |
| $2\circ 77$ | $B$ | ((7-2)-(•−)) | 5-1 |
| $2\circ 77$ | $W$ | (2-(•−)) | 6-1 |
| $1\circ 1\overline{3}$ | $B$ | ((1-$\overline{3}$)-1) | 3-6 |
| $1\circ 1\overline{3}$ | $W$ | ((1-$\overline{3}$)-($\overline{3}$-1)) | 3-5 |
| $1\circ\overline{4}7$ | $W$ | ((1-$\overline{4}$)-($\overline{4}$-7)) | 3-5 |
| $\overline{5}7\circ 7$ | $B$ | (7-($\overline{5}$-7) | 2-5 |
| $\overline{5}7\circ 7$ | $W$ | (7-$\overline{5}$)-($\overline{5}$-7) | 3-5 |
| $11\circ 6$ | $B$ | ((−•)-(6-1)) | 7-3 |
| $11\circ 6$ | $W$ | ((−•)-6) | 7-2 |

| tile | *st* | edges | relative |
|---|---|---|---|
| $11\circ 2$ | $B$ | ((−•)-(2-1)) | 7-3 |
| $11\circ 2$ | $W$ | ((−•)-2) | 7-2 |
| $37\circ 7$ | $B$ | (7-(3-7)) | 2-5 |
| $37\circ 7$ | $W$ | ((7-3)-(3-7)) | 3-5 |
| $1\circ 14$ | $W$ | ((1-4)-(4-1)) | 3-5 |
| $\overline{5}\circ 77$ | $B$ | ((7-$\overline{5}$)-(•−)) | 5-1 |
| $\overline{66}7\circ$ | $W$ | ($\overline{6}$-$\overline{6}$) | 2-4 |
| $1\overline{22}\circ$ | $W$ | ($\overline{2}$-$\overline{2}$) | 4-6 |
| $11\circ\overline{3}$ | $B$ | ((−•)-($\overline{3}$-1)) | 7-3 |
| $47\circ 7$ | $W$ | ((7-4)-(4-7)) | 3-5 |
| $1\circ 15$ | $B$ | ((1-5)-1) | 3-6 |
| $1\circ 15$ | $W$ | ((1-5)-(5-1) | 3-5 |
| $6\circ 77$ | $B$ | ((7-6)-(•−)) | 5-1 |
| $6\circ 77$ | $W$ | (6-(•−)) | 6-1 |
| $137\circ$ | $B$ | ((1-3)-7) | 3-6 |
| $157\circ$ | $B$ | (1-(5-7)) | 2-5 |

**Table 2** *Table of the centres and petals equipped with the isoclines.*

In the sequel, the isoclines 0, 5, 10 and 15, which we already called the rows of the mantilla in the section 3.1, will also be called **active** as they contain the seeds. The other isoclines play a passive rôle. Accordingly, the isoclines 1 to4, 6 to 10, 11 to 15 and 16 to 19 can be seen as behaving exactly in the same way as the passive rows of section 3.4. They only convey the vertical signals, the implementation of which is investigated in the next subsection. For this reason, the just indicated groups of isoclines will be called the **passive zones**.

It is also not difficult to append marks to the numbering in such a way that the active rows are distinguished from the passive zones. See [10] for a possible solution.

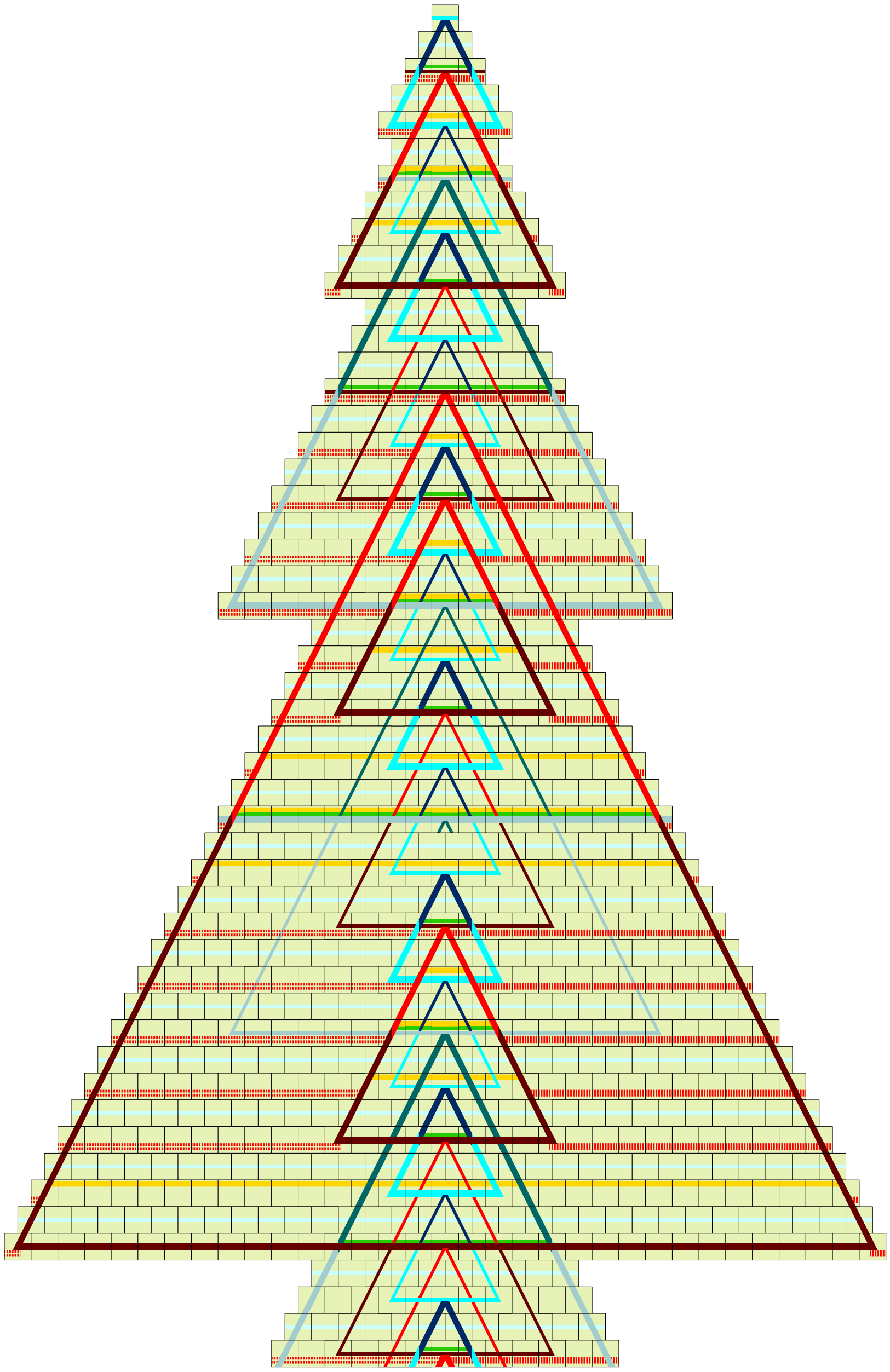

**Figure 11** *Triangles and phantoms up to the generation* 3. *This invovles only the tiles indicated by formulas* $(B)$, $(V)$, $(M)$, $(C)$, $(L_u\varphi)$, $(L_ut)$, $(L_\ell)$ *and the tiles*

*of figure* 8*. Note the vertex of the phantom of the generation* 3 *at the mid-point of the basis of the triangle* $R$ *of the generation* 3*. Also note the part of the basis of another phantom of the generation* 3 *around the vertex of* $R$*.*

## 3.2 Implementation of interwoven triangles in the hyperbolic plane

The set of tiles which we described in section 3.4 is not, strictly speaking a set of Wang tiles.

However, its implementation in the heptagrid will consist of true *à la* Wang tiles. But for this purpose, we have to first indicate how we implement the trilaterals in the hyperbolic plane.

### 3.2.1 The trilaterals in the hyperbolic plane

In section 4.1, we already implemented an **isocline signal** which defines the isoclines of section 3.1. The rows which we have just defined among them in the prevISOus sub-section are the rows along which we shall construct the bases of the implementation of the trilaterals. These rows will also convey the horizontal signals needed by our construction.

Now, that we know what are the rows, we have to indicate what are the verticals, what is the axis and how to define the legs of the triangles as the notion of slope is meaningless in the hyperbolic plane.

The legs of the trilaterals will be implemented along the border of the trees of the mantilla rooted at a seed, where the vertex stands. The left-hand side leg goes on the leftmost branch and the right-hand side leg along the rightmost branch. The corners are the intersections of the legs with an isocline which will also support the basis of the trilateral.

Note that the implementation of the Euclidean trilaterals are not triangles of the hyperbolic plane. However, we shall keep the same terms as in the Euclidean case. First, it raises no confusion and, second, it is simpler.

Now, from this definition of the implementation, a lot of questions arise at the level of signals inside and between trilaterals.

The first question is about the verticals and the axis: what are here their implementation?

We may consider the borders of a tree of the mantilla as verticals. In the part devoted to the simulation of a Turing machine, we shall indicate what are the actual verticals but, up to now, we do not need them. The axis is a more urgent question. In fact there is no axis as we had in the Euclidean case. However, the axis has an important rôle in the Euclidean case, as it triggers the positionment of the vertices of the trilaterals. In the hyperbolic case, there is no possible axis and we have to replace it by a procedure which gives the same result. Now, from lemma 3, we can apply the following: each active seed of an isocline 0 dispatches a signal to its sons and grandchildren until it reaches the next isocline 0. Call this signal the **scent**. When the signal reaches an isocline 5, 10 or 15: if it meets a seed, the seed becomes active and the signal goes down further. If it meets a tile which is not a seed, then it stops. Accordingly, not all seeds of an isocline 0 is reached by the scent from a seed of the just upper isocline 15. Note that the property of an active seed to diffuse the scent is not attached to the status nor the colour of the trilateral emitted by this seed.

This mechanism allows to trigger the diffusion of the green signal on the isoclines 15. It is important to remember that, starting from the second generation, any vertex of a trilateral appears within a blue-0 phantom. Accordingly, the starting of all green signals is performed by the phantoms of the generation 0.

From this, it is natural to define the following:

*A* **branch** *is a sequence of active seeds* $\{\sigma_i\}_{i\in\mathbb{N}}$*, either indexed by* $\mathbb{N}$ *or by* $\mathbb{Z}$*, such that the tree of the mantilla rooted at* $\sigma_i$ *contains the tree of the mantilla rooted at* $\sigma_{i+1}$ *and such that the distance between* $\sigma_i$ *and* $\sigma_{i+1}$ *in isoclines of the mantilla is* 5.

Note that a branch starting from a vertex of a blue-0 triangle is infinite and that the scent exactly marks all the branches.

If the mantilla possesses an ultra-thread, there is a branch indexed by $\mathbb{Z}$ which we shall call a **traversal**. If this is not the case, all branches are indexed by $\mathbb{N}$. However, in the case of an ultra-thread, there are also branches only indexed by $\mathbb{N}$: the condition on the distance 5 may be not satisfied between an active seed $\sigma$ and another one $\xi$ belonging to the tree rooted at $\sigma$ and such that there is no intermediate tree of the mantilla rooted at an active seed between $\sigma$ and $\xi$. As an example, computer checkings show that in a tree of the mantilla rooted at a seed, say on an isocline 0, there are 6 seeds on the isocline 5 which are inside the tree and 324 seeds on the isocline 10 which are inside the tree. Note that the number of nodes inside the tree on the isocline 5 is 144 and that there are 17,711 nodes inside the tree on the isocline 10. Now, if any seed on the isocline 10 would be a grand-children of a seed on the isocline 5, there would be at most 36 of them. Consequently, the phenomenon which we have described happens very often.

Now, in any case, the trilaterals which are on a branch implement a semi-infinite model of the abstract brackets. We know that such a model is obtained as a cut of an infinite model. Our first condition will be that in each realization of the tiling, the branches implement cuts of the same infinite model. Of course, in the case of an ultra-thread, the traversal will implement the same infinite model.

Clearly, this requires to **synchronize** the choices of the triangles at each generation on different branches as these branches can be independent in a mantilla without ultra-thread. The synchronization is also needed in all the cases as we do not know a priori, whether the realization possesses or not an ultra-thread.

As already mentioned, the basis of a trilateral of the generation 0 is very large. As there are 36 active seeds on it, this means that a single triangle $B$ of the generation 0 defines 36 phantoms whose vertex is on the basis of the triangle. Now, consider that the isocline 5 which crosses $B$ gives rise to red triangles. There are 6 of them. And so, we have to define how to manage the horizontal red signals emanating from these triangles. We also have that all these triangles cut the basis of $B$.

In order to handle more easily the problem, define the **latitude** of a trilateral as the set of rows of the mantilla from its vertex downto its basis, both included. The length of this set is called the **amplitude** of the zone constituted by the latitude. Of course, trilaterals of a same generation have the same amplitude. The synchronization has, as a consequence that the latitudes of triangles of the same generation from one branch to another exactly coincide or do not overlap. Consequently, the same also holds for trilaterals.

This induces a problem for the propagation of horizontal red signals. We have defined them with a laterality: left-hand or right-hand side. Up to now, there is a tile which allows to join one laterality with the other: it is the tile of the vertex of a red triangle. But here, the opposite meeting occurs between different laterality, *i.e.* a left-hand side signal coming from the right meets a right-hand side signal coming from the left. For this purpose, we shall just introduce a new tile illustrated in the section 3.4: its left-hand side half is a red-hand side horizontal red signal and its right-hand side half is a left-hand side horizontal red signal. Call it the **red join tile**.

Coming back to the realization of semi-infinite models, this entails a similar situation connected with missing trilaterals. Remember that, in a semi-infinite model, all active intervals which contain the point where the cut is performed are removed from the model. Depending on the infinite model, this leads to remove either finitely many active intervals or infinitely many of them. In our implementation, we first remember that a seed $\sigma$ always defines another seed inside the tree of the mantilla rooted at $\sigma$, at a distance 5 from $\sigma$. This means that once a vertex of a trilateral is present, the whole trilateral is present and, consequently, this is also the case for all its followers, below, of the same generation. However, as the considered seed also generates a basis, it may happen that there is no seed to realize a vertex containing this point of its basis. The basis is still propagated by the seed which cannot foresee this possibility. In some sense, this basis is a lost signal and so, we also have to handle this phenomenon and to prove that this will not disturb the process we used up to now to construct the trilaterals.

*Managing lost signals and syncronization*

The implementation of the interwoven triangles in the hyperbolic plane entails a new ingredient to our tiling. In the Euclidean situation, we have horizontal red signals. In the hyperbolic case, we also have blue horizontal signals. This is due to the following problem which did not occur in the Euclidean case as, inside a triangle exactly one trilateral is generated by the mid-distance line of the triangle.

In the hyperbolic case, inside a triangle of a given generation, we have several trilaterals generated on the mid-distance row of this triangle. This is clear from what we have already noticed. But also, within the same latitude, we have infinitely many trilaterals, even if on the branches which are raised within this latitude, there may be missing trilaterals in this latitude, as stated by the following statement.

**Lemma 19** *Within a given latitude corresponding to an interval of the model realized by the tiling, there are infinitely many trilaterals.*

The easy proof is given in [10].

Due to the multiplicity of the trilaterals of the same generation within a given latitude and also due to the occurrence of a lot of bases which cannot find their vertex, it is important to be sure that a leg will distinguish between a lost basis and the basis with which it will make a corner of the trilateral under construction.

Let us look at the case of a lost basis $\beta$ of a red triangle of the generation $m$. It may happen that such a basis occurs not far from a leg of a bigger red triangle, *i.e.* of the generation $n$ with $n \geq m+2$. As a basis of a red triangle whose corner is not yet installed and as phantoms do not emit signals, the basis has no horizontal red signal with it. And so, it seems that there is no difference for this leg of the generation $n$ between this basis and the expected basis of the generation $n$. In fact, there is a difference. The lost basis $\beta$ is of the generation $m$ with $m < n$. Now, on the latitude of the lost basis, there are infinitely many other bases which are not lost, as proved by lemma 19. And two of them, the closest one to the left and the closest one to the right emit horizontal red signals starting from their corner which are closest to $\beta$. As there is no leg to stop this signal, $\beta$ is accompanied by such a signal. This is not the case for a basis of the generation $n$: at the time of its construction, the red signal is not installed at the level of the basis. This gives us a way to distinguish between the two situations.

First, we have to remark that there is still a problem with the basis of a red phantom. Such a basis is accompanied by the horizontal red signal of a vertex of a red triangle as a basis of a trilateral, inside the triangle which has generated it, which, to its turn, generates trilaterals of the other kind. Now, it seems that there is no difference between a lost basis of a red phantom of the generation $m$ and a basis of a red phantom the generation $n$ which looks after the legs of its phantom. In fact, there is a difference. In the tiles which we devised in section 3.4, the horizontal signal is below the basis. It is enough to fix a level in the tile at which the normal horizontal signal travels. We shall call such a signal the **upper** one as we agree to place it above the basis signal on the tile. Then, in the case of a basis of a red phantom, we shall agree that the signal emitted by the vertex of a triangle generated by this basis is **below** the basis. Accordingly, we shall say that it is the **lower** signal. By construction, the lower and upper signals have the same colour as the basis. Their laterality is fixed by elements which give them rise: the corner, for the upper signal, the vertex for the lower. It is plain that between two consecutive elements, both signals can be joined as, in both cases, the right left-hand side signal comes from the left and the left-hand side signal comes from the right. Consequently, a basis of a red phantom of the generation $m$ will have two horizontal red signals outside the phantom: the upper signal generated by corners of the neighbouring phantoms of the same latitude and the lower signal which is below the basis and which is generated by the vertices of triangles occurring on this row of the mantilla. On the other hand, inside the angle defined by the legs of a red phantom of the generation $n$, the expected basis has a single signal, the lower one and so, the legs of the phantom may notice the difference.

From this, we immediately remark that for blue trilaterals, in the present state of the tiles, there is a problem: as they do not emit signals, there is no way to distinguish between a basis which did not find legs and an expected one of a later generation. But the just above analysis points at a solution: let us decide that blue triangles, both blue 0 and simple blue ones will emit horizontal blue signals outside themselves. The signals will be managed exactly as the red ones. We shall have a left-hand side and a right-hand side blue horizontal signal. We have a tile to join signals between triangles lying in the same latitude which we call the **blue join tile**. Also for bases of blue phantoms, the signal induced by the vertices of blue triangles generated by this basis travels below the basis and the upper signal, generated by corners travels at the standard level of the horizontal blue signal, above the basis. In the tiles, this standard level is the same as the standard level of the red signals. As we have dotted signals, the superposition of an upper horizontal blue signal and an upper horizontal red signal on the same row will raise no problem. Note, that this superposition occurs very often as it is the general case, see figure 12.

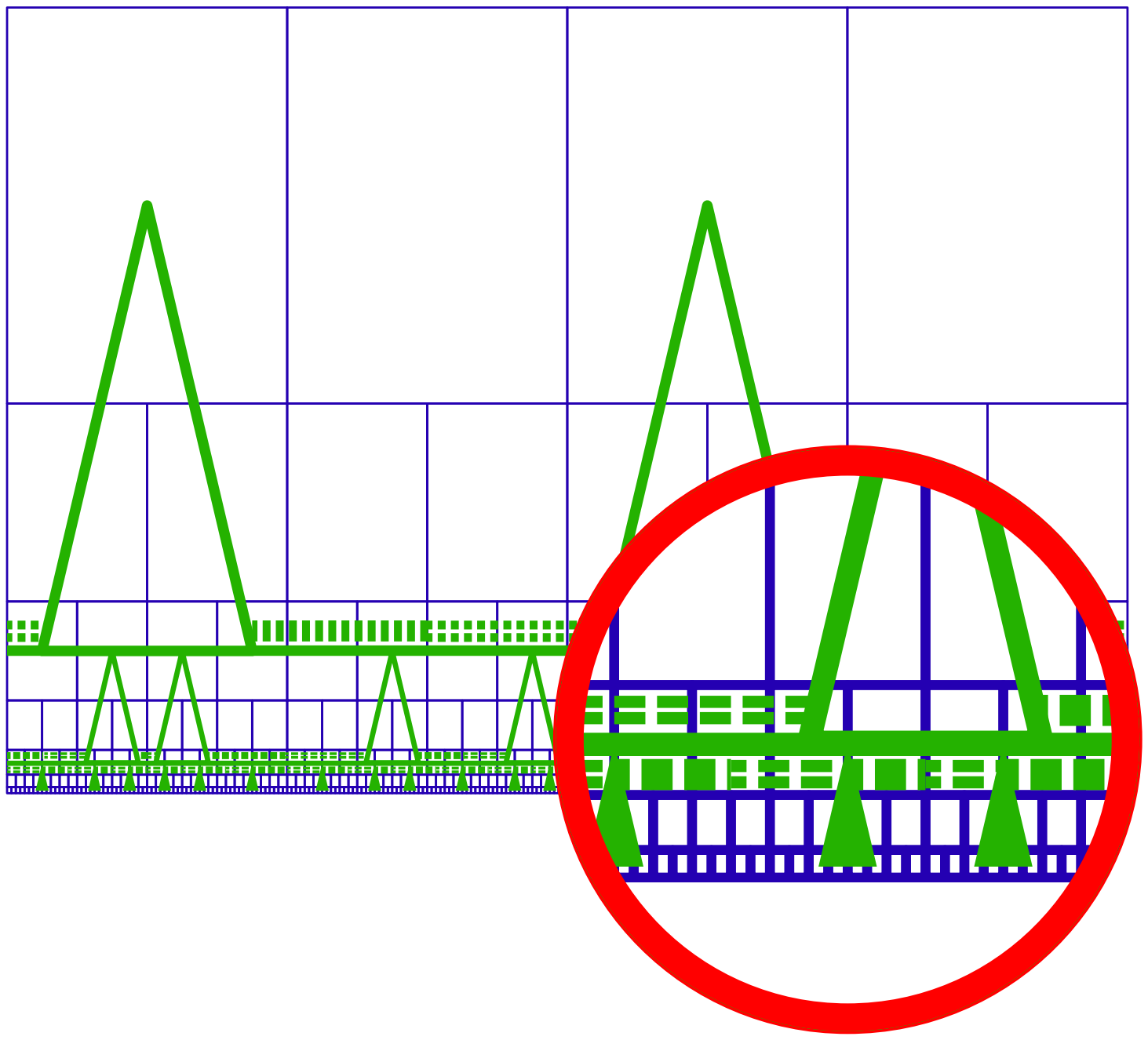

**Figure 12** *Management of the bases as a synchronization signal fro trilaterals of the same latitude.*

And now, there is a good news: lost signals are not that lost. As they travel on the same row as all trilaterals of a given latitude, they may be interpreted as a synchronization signal. Moreover, there is no need to stop them when they meet a corner and we decide that all bases of the same isocline merge. The difference between inside and outside the trilateral is clearly made by the presence or not of an upper horizontal signal of the same colour: if the upper signal is present, this part of the basis is outside the trilateral, if not, this part is inside the trilateral. As suggested by figure 12, the basis which now runs across the whole hyperbolic plane on its isocline, from the left to the right, can be split into intervals of two kinds. In one kind, all the intervals are inside a trilateral. In the other kind, all the intervals are outside any trilateral. Of course, we only consider maximal intervalls with this property. Note that this notion of inside and outside is specific to the generation of the basis. Accordingly, an outside interval of a basis may fall inside a trilateral of another generation. In our sequel, we shall speak of an **open** basis when we consider an interval which is inside a trilateral of the generation of the basis and we shall speak of a **covered** basis when it is outside. Clearly, an open basis has no upper horizontal signal of the same colour while a covered basis is necessarily accompanied by such a signal.

On an outside interval of a basis, there are indeed two hoizontal signals of the same colour as the basis. One is a right-hand side signal coming from the left and the other is a left-hand side signal coming from the right. We know that they are joined by an appropriate. Note that such a junction is unique. A similar remark can be made for lower signals. They also merge, being joined by a similar tile when the opposite lateralities come from directions which are opposite to the sense they indicate, and they are joined by a vertex of a triangle when the opposite lateralities come from the sense which they indicate.

At the level of the tiles, the upper horizontal signal of the covered basis merges with the upper horizontal signal of the corner of the same laterality and the same colour.

Repeat that these tiles allow to a right-hand side horizontal signal coming from the left with a left-hand side signal coming from the right. Note that such a tile cannot be used inside a triangle $T$: the opposite junction is only made by the presence of another triangle of the same colour inside $T$. There is no other possibility as the mantilla and the isoclines prevent to rotate the tiles. And so, appending the above joining tiles does not affect the way we use the horizontal signals inside a triangle.

This new situation fixes what happens in between consecutive trilaterals of the same latitude. It is illustrated by figure 12. In our sequel, we shall only speak of bases.

As we introduced several important changes, we have to pay a new visit to the algorithm. Just after, we shall complete the set of tiles.

### 3.2.2 The algorithm revisited

First, we construct a realization of the mantilla, see [9].

Then, we draw the isoclines, taking at random an $F$-centre which is the son of a $G$-flower. While drawing the isoclines, we assign them numbers in 0..19, periodically. By definition, the numbering increases, downwards.

*The step* 0

*Triangles*

The step 0 of the construction is the drawing of the blue-0 trilaterals and its accompanying signals of the generation 0: the blue signals, to the left and to the right emitted by the legs of the triangles of this generation.

For this purpose, each seed on the isocline 0 is the vertex of a blue-0 triangle whose legs are spread until the next iscoline 10. The seed also sends a signal inside the triangle, the **scent** in the conditions we explained in sub-section 4.2.1. As this process is achieved when the construction of the phantoms of the generation 0 is completed, we consider it as independent of the further generations and, accordingly we shall not mention it. It will be implicit when we shall say that the active seeds of a certain generation do this and that. Each seed of the isocline 0 also emits three horizontal signals. One is the basis of a phantom of the generation 0 which is drawn along the whole isocline, from left to right, across the whole hyperbolic plane. The other signals are the horizontal blue signals emitted by the triangle. Due to the presence of the basis, they travel below the basis: they are the lower signals. They meet other lower signals emitted by the neighbouring seeds of the isocline 0 thanks to the blue join tile. And this pehnomenon is repeated along the whole isocline.

The scent raises the green line when it reaches the isocline 5 and also, later, when it reaches the isocline 15. This green line is inside the triangle and it grows until it reaches the leg of the triangle where it is stopped. All seeds which are on the green line and which receive the scent become active. They emit legs of a red trilateral. We choose at random one isocline 5 in order to fix those which will generate the triangles of the generation 1. The blue 0 legs of the generation 0 go on until they reach the isocline 10. The seeds of the isocline 5 which are inside a blue-0 triangle are active and they also emit legs of the corresponding trilateral. They emit the basis of red phantoms which will travel on the isocline 5, to both its ends at infinity. They accompany this basis, by horizontal red signals which are placed below the basis. The seeds also similarly emit a scent inside the triangle which reaches the isocline 10. The seeds of the isocline 10 which receive the scent, and only them, become active. They emit legs of the phantoms of the generation 0. They also emit the basis of the blue-0 triangle. The basis signals merge and when they reach the legs of the seed, they form a corner at their intersection. The basis signal travels along the isocline in both directions, until its ends.

Note that the basis meets the legs of red triangles before it meets its expected legs. As there is no red triangle of a previous generation, there cannot be red signals inside a triangle of the generation 1 and so, the legs of the triangle determine a yellow signal on the isocline 10.

Besides the red basis, there is a single horizontal signal inside a triangle of the generation 0: the green signal which is on the isocline 5. The triangle emits six horizontal blue signals: three on its left-hand side and three on its right-hand side.

Between triangles within the same latitude, these signals meet, using the blue join tile. We may place it whereever we wish, vertices being excepted. Remeber that between two consecutive corners of distinct triangles, there can only be one blue join tile. Indeed, the seeds of the isocline 10 are active if and only if they are on a basis of a blue-0 triangle and they receive the scent of a red triangle generated at the isocline 5.

*Phantoms*

Now, let us look at the construction of the phantoms which are raised by the active seeds of the isoclines 10. Their latitudes are defined by the isoclines 10 up to 19 and the next isocline 0.

Consider an active seed $\sigma$ of the isocline 10. It emits legs of a phantom and it also emits the basis of a blue-0 triangle. Now, as $\sigma$ is an active seed of the isocline 10, it is on a branch starting from a seed $\sigma_0$ of the isocline 0 belonging to the same interval [0..19] of isoclines as the considered isocline 10. Accordingly, the tree of the mantilla issued from $\sigma$ is contained in the tree of the mantilla issued from $\sigma_0$. Accordingly, the legs issued from $\sigma$ must meet a basis on the next isocline 0, downwards.

The legs of the phantom go down until they reach the isocline 15. The seed $\sigma$ also diffuses a scent inside the phantom. When it reaches the isocline 15 it stops on tiles which are not a seed. When it meets a seed, as in the case of a triangle, the seed becomes active and the scent triggers the construction of a green line inside the phantom. But, contrarily to what happens in a triangle, in the case of a phantom, the legs do not stop the green line which goes on outside the legs of the phantom, looking after legs of a triangle which will stop the signal.

In our situation of the step 0, the different green lines which we have inside the legs of each red triangle of the generation 1 meet and, at the same time, they merge. Soon, they meet the legs of each triangle which stop them. At the same time, inside each red triangle, as long as the green line meets a seed which also receives the scent emitted by a the vertex of a phantom, the seed becomes active.

Now, at this moment, it is the starting point of the step 1.

But before looking at this step, we have to see what happens with the bases triggered by the seeds which lie on the isocline 0.

The basis itself goes to infinity in both directions and, consequently, it will serve as a basis for all blue-0 phantom of this latitude. Now, the basis is accompanied by a horizontal blue signal below the basis, a left-hand side signal to the left of the seed, a right-hand side one to its right.

For these signals, either the basis meet legs of a blue-0 phantom, or they do not. If both signals starting from the seed meet such legs, each one on its side, being inside the blue-0 phantom, they form a corner with the legs, completing the construction of the corresponding blue 0 phantom. We notice that there is no horizontal signal above the basis inside the phantom. Also, at the joining point with the legs, the horizontal signal which is below goes on: as all seeds of the isocline 0 become active, a lot of them which are outside the basis of any phantom will emit the legs of a triangle and, consequently, below the basis of the phantom which we consider, the lower horizontal blue signals. They will join the one we consider thanks to the blue join tiles. Infinitely often, the bases of the isoclines 0 meet the legs of a blue-0 phantom inside the angle determined by the legs and this way, a phantom is constituted. But also infinitely often, much more on any interval of such an isocline than bases which succeed, the basis does not meet such legs. Note that this is known by the seeds of the isocline 0. If a seed of the isocline 0 receives the scent, it knows that it belongs to a branch which starts at least from the previous isocline 0. The branch may even join a traversal, but the seed cannot know it. But if the seed of the isocline 0 receives no scent, it knows that it starts a new branch.

A covered basis goes on on each side of such a corner until it meets a corner, which must happen, as infinitely many phantoms do exist within the considered latitude. When a corner is met, the upper horizontal signal merges with that of the corner and, in the next sub-section, we look at the tiles which implement this junction. Now, as the corners of the triangles emit upper horizontal blue signals, these signals will pass over the bases of the generation 0 which do not find legs to form a blue-0 triangles. The tiles will also bear this indication.

We shall deal with the other bases, those of the blue-0 phantoms and those of the red trilaterals of the generation 1 in the next point devoted to the induction step.

*The step* $n+1$

We define the starting point of the step $n+1$ by the time when, drawing the legs of a triangle of the generation $2n+1$, they reach their mid-distance row. The first thing we do is to decide which of the concerned isocline 15 becomes a row of vertices of the simple blue triangles of the generation $2n+2$. Note that in this description, we shall never indicate that legs of a triangle emit upper horizontal signals of their laterality and in their colour, outside the triangle. This will be considered as obvious.

As the induction hypothesis, we assume the following:

(*i*) All trilaterals of the generation $m$ with $m < 2n$ are complete. In particular, all the horizontal signals which they emit are drawn. A signal can go out from a triangle of its colour. It enters a triangle $T$ of its colour if and only if there is a leg of a triangle of its colour inside $T$ and on this row. A signal crosses the triangles of the other colour which intersect its latitude and it also crosses any phantom which intersects its latitude. However, for trilaterals of their colour, the legs require to be crossed by signals of their laterality. For horizontal signals of another colour than that of the crossed legs, there is no restriction.
(*ii*) The bases for the generations $m < 2n$ are completely drawn, up to infinity. The upper horizontal of the same colour accompanies them between consecutive trilaterals within the same latitude. It has the lateralities of the corners which emit them and the opposite lateralities are joined. For phantoms, the lower horizontal signal is everywhere present. Its different lateralities are joined appropriately.
(*iii*) All triangles of the generation $2n$ are completed, with the same meaning as in (*i*).
(*iv*) All bases of the generation $2n$ are completely drawn up to infinity. They are covered as required by an upper horizontal blue signal between consecutive trilaterals of the same latitude.
(*v*) All the vertices of the phantoms of the generation $2n$ are determined and the legs which they emit are drawn until they meet the first green line of an isocline 15.
(*vi*) All the active seeds for the triangles of the generation $2n+1$ are determined and the legs are drawn until they reach the mid-distance row determined by the phantoms of the generation $2n$: the isocline 15 found at (*v*).

Note that for $n = 0$, this is exactly the point which we have reached after the step 0.

Now, as the mid-point of the legs of the phantoms of the generation $2n$ as well as those of the triangles of the generation $2n+1$ are determined, we go on drawing these lines until they meet their respective bases.

For the legs of the phantom, we are now in their second half, they go on as long as they cross upper horizontal blue signals, whatever the accompanying signal. In particular, the upper horizontal blue signal may be accompanied by a red one.

The bases of the phantoms are emitted by the seeds which are on the mid-distance row of red triangles of the generation $2n-1$, this row being the closest to the row $\rho$ of the vertices of the considered phantoms, and below $\rho$. The bases grow, accompanied by a lower horizontal blue signal, and they travel on their row until they meet the legs. When this happens, if the basis is inside the angle determined by the legs, it cannot be covered by an upper horizontal blue signal. Indeed, a basis of a given generation does not travel on a row which belongs to the latitude of a triangle of its colour and of a previous generation. A horizontal signal of the generation $2n$ on this row necessarily comes from the vertex of a triangle. Now, the signal can be an upper one only if it was put on this level by the corner of some phantom of the generation $2n$. This means that it has met the corresponding legs. Now, if this happens for one phantom, this also happens for the others, and at the same time.

Note that there is not a confusion with a basis which is outside a blue phantom of a previous generation: such a basis has two horizontal blue signals: an upper and a lower one. Now, the upper signal is emitted by a triangle which is inside the phantom. The row at which travels the expected basis is the first one, inside the angle determined by the phantom, on which there is no upper horizontal blue signal. Accordingly, the leg meets its basis on this row, as indicated above. In the next section, we shall see how the tiles force the right choice, thanks to the laterality of the signals, see page 0 in the proof of lemma 20.

This also fixes the row of the basis outside the phantoms of the generation $2n$. As indicated before, the basis signal $\beta$ is drawn along the isocline from left to right across the whole hyperbolic plane. As there are seeds on an isocline 0 at any distance in numer of isoclines from a given row and as far as desired from a given point on a row, there are always legs of phantoms of the generation $2n$ which meet $\beta$. Now, the corner of these legs emit upper horizontal blue signals in the direction of their laterality. Consequently, between two consecutive corners belonging to different phantoms, the upper signal go towards each other. The blue join tile allows to perform the meet-ing. Now, $\beta$ also crosses seeds which are not inside a phantom. Accordingly, these seeds emit a lower horizontal blue signal. Between two consecutive vertices, the configuration of the signals is the same as the configuration of an upper signal between consecutive corners of different phantoms. And so, the same joining pattern as for the upper signal allows to join two lower signals meeting each other. The exact configuration of the corresponding tiles is given in the next section.

At this point, the trialterals of the generation $2n$ are completed, including the signals which accompany their bases along the whole row.

Let us now look at what happens with the legs of the triangles of the generation $2n+1$.

They do the same as the legs of the phantoms of the generation $2n$. This time they look at the upper horizontal red signals which are interpreted in the same way. However, there is a difference: the legs also mark the free rows inside a triangle with the yellow signal. They perform this task exactly as it is depicted in the Euclidean case. Accordingly, the legs find their bases as the corresponding bases are emitted by the active seeds of the concerned isocline 15. The bases are drawn along the whole isocline, running without interruption from left to right, across the whole hyperbolic plane. Between two triangles of this generation and of the same latitude, the accompanying horizontal red signals make use of the same mechanism as what was just described for the phantom of the generation $2n$. Note that here, there is no lower signal as vertices of phantoms do not emit signals.

Accordingly, the red triangles of the generation $2n+1$ are completed.

Now, we know that the seeds which are on the basis of a triangle of the generation $2n+1$ and which are also on a green line of a triangle of the generation $2n$ and which receive the scent of a seed of a blue-0 phantom are active and only them. We also know that they emit legs of phantoms of the generation $2n+1$, the issue of the basis of triangles which they emit being already considered.

The legs take into account the horizontal red signals only which emanate from the triangles of the previous generations. We know that in any model of the abstract brackets, a silent interval has the same structure of the intervals it contains as an active one. Accordingly, things happen exactly as it does for the triangles of this generation. The green signal is met at the place where the green line of the phantoms of the generation $2n$ is. Indeed, this is the mid-distance row for these phantoms. Now, we know that for phantoms, the green signal is not stopped by the legs. Accordingly the signal goes further on both its sides. As this row is free of signals of the previous generations and as it does not contain a lateral signal of the generation $2n$, it is also free for the generation $2n+1$. Now, the legs of the phantoms of the generation $2n+1$ also meet this green signal which also crosses their legs. The legs go on again until they meet the bases issued by the next free row, this time free of upper horizontal red signal. Indeed, this time the legs reach a row which is the mid-distance row of a triangle of the generation $2n$. We know that the active seeds inside those triangles in this row emit bases of phantoms. The discussion is exactly the same as for the phantoms of the generation $2n$.

And so, $(i)$ and $(ii)$ are true for the step $n+1$.

Now, while the phantom of the generation $2n$ and the triangles of the generation $2n+1$ achieve their completion, the seeds on the defined rows for the vertices of the triangles of the generation $2n+2$ have started to emit signals: they are the seeds which are on the just selected isoclines 15 and which receive the scent of a blue-0 phantom and which are on the green line. Accordingly, these seeds are also on a branch.

Let $T$ be a triangle of the generation $2n+2$. Its legs go downwards as long as they meet an upper horizontal blue signal. One of these signals is accompanied by the basis of the triangle $T_0$ of the generation $2n+1$ which has generated $T$. During their travel, the legs of $T$ also meet bases of phantoms from the generation 0 until the generation $2n$. For what is the blue phantoms, these bases are emitted by the vertices of the blue triangles already present in $T$. All these phantoms contain the vertex of $T$. Accordingly, as their bases, at this point, did not yet reached their legs, they are open and they are onlu accompanied by the lower signal only for what is their colour. And so, a blue basis is not accompanied by the upper horizontal blue signal. The leg of $T$ must not disturb this configuration by introducing an upper horizontal blue signal outside. Consequently, the leg does not emit such a signal in this case. It considers that the lower signal is enough to indicate that the considered row is not free. Note that these signals occur before the legs of $T$ meet the basis of the triangle $T_0$.

Now, there are other basis signal of a phantom which are accompanied by an upper horizontal signal: they concern phantoms which are inside $T$. The legs also must not undo such signal. We see, in the next section, in the proof of lemma 20, that the tiles make the difference and do not allow to replace the established situation by the opposite one. For each phantom of such a generation, from the generation 0 until the generation $2n$, its basis accompanies the upper horizontal blue signal. Note that these signals occur along the whole way of the legs of $T$.

At last, the legs of $T$ meet the first free row. It is the green signal already met by the phantoms which are generated by the appropriate seeds of the basis of $T_0$. Consider $P$ such a phantom. As the green signal which runs on the mid-distance row of $P$ is not stopped by the legs of $P$, the signal will meet the legs of $T$ and it will be stopped by these legs. Indeed, the legs go until this row as, from lemma 9, the inside of the half of a blue active interval exactly consists of juxtaposition of the blue active intervals of the previous generations. And so, the free row precisely occurs at this row.

The legs go on, it is now their second part which reacts differently than the fist part as some events which happen for the first part cannot happen for them. Again, the legs meet upper horizontal blue signals from the triangles of the previous generations. Then, we have the basis emitted by the green signal of the triangles of the generation $2n+1$ which are issued from $P$. This basis meets also similar bases produced by red triangles of the generation $2n+1$, all contained in the tree of the mantilla which is rooted at the vertex of $T$. Accordingly, the basis will meet the legs. As previously, the basis travels on a row which has no horizontal signal. Accordingly, it is clearly identified and it cannot be confused with previous blue bases which are accompanied by a upper horizontal blue signal. We again have the same discussion for what happens with the bases outside the triangle: the corners emit an upper horizontal blue signal which meets a similar signal but of the opposite laterality emitted by the neighbouring corner of another triangle. The blue join tile allows the signal to meet.

Accordingly, the items $(iii)$ and $(iv)$ are also true at the step $n+1$.

Now, the bases of the triangles of the generation $2n+2$ contain active seeds which emitted them and which now emit the legs of the phantoms of the generation $2n+2$. Let $P_2$ be such a phantom. Its legs go down and thanks to lemma 9, we know that the first occurrence of a free green signal is on the mid-distance row of the phantoms of the generation $2n+1$ which are inside the triangle $T_1$ of the generation $2n+1$ which has generated $P_2$. And, as legs of a phantom, the legs of $P_2$ do not stop the green signal.

Accordingly, the item $(v)$ of the induction hypothesis is also true for the step $n+1$.

Now, when the triangles of the generation $2n+2$ have reached their mid-distance row, this has triggered the construction of the trilaterals of the generation $2n+3$. At this time, one of the concerned isocline 15 is chosen as a row where all the active seeds generate a triangle of the generation $2n+3$. The construction of the legs go as usual, the same phenomena as for the construction of the triangles of the generation $2n+2$ occur for what regards the crossing of bases of the phantoms of the generations from 0 up to $2n+1$ which may be accompanied or not by an upper horizontal signal. As in the case of the generation $2n+2$, the legs of the triangles of the generation $2n+3$ do not undo the previous constructions. Now, as red legs, the legs of the triangles of the generation $2n+3$ detect the free rows inside the triangle by the technique already described for the Euclidean triangles. And the first green signal met by the legs is the one we just describe for the phantom $P_2$ or one of its copies in the same latitude.

And this proves that the item $(vi)$ is also true for the step $n+1$.

Consequently, all items of the induction hypothesis are also true for the step $n+1$.

This proves that the construction of a tiling implementing the interwoven triangles is possible by using the just described algorithm.

### 3.2.3 The new set of tiles

The implementation of the square tiles of section 3.4.2 in heptogonal tiles does not present major difficulties. It is simply a translation process which we now indicate.

*From the square tiles to the heptagonal ones*

The square tiles of section 3.4 bear two kinds of signals: vertical and horizontal ones.

In the translation of the square tiles into the heptagonal ones, the horizontal signals will follow the route of the isocline signal. For the verticals, the things are not more complex. Going back to the construction of the isoclines and remembering lemma 2, we first notice that we need only to use black nodes for the vertical signals. The route of a left-hand side leg goes along the leftmost branch of a tree of the mantilla. Accordingly, in our black tiles, this route goes from the edge 1 to the edge 4 in paternal coordinates. The route for a right-hand side leg goes along the righmost branch of a tree of the mantilla. And so, this route goes from the edge 2 to the edge 6 in paternal coordinates: remember that the right-hand side border of a tree of the mantilla consists of the black-nodes whose father is not in this tree.

For this reason, we do not provide the representation of these tiles in their heptagonal setting. Also, we shall keep the Euclidean format to present the new tiles required for the horizontal signals, to manage the new definition of the bases connected with the synchronization.

*The tiles for the horizontal signals*

We mainly have two kinds of new tiles: the tiles for introducing the horizontal blue signals in our previous setting, *i.e.* for inside the trilaterals and the tiles for joining signals outside the trilaterals, *i.e.* between two consecutive trilaterals of the same latitude.

As already mentioned, we shall not list pictures of the tiles: there are two many of them. We shall use the technique introduced in the Euclidean model. We represent the tiles by formulas given by tables 3, 4 and 5. The formulas are established as this was the case in the Euclidean setting. They are simply more complex, and the number of patterns is bigger. We go from 192 patterns to 1,227 ones, which is a bit more than six times the number of the Euclidean situation.

There is some difference in the elementary patterns. We introduce new patterns in order to implement the new signals which we introduced in this section. Below, we indicate which patterns are kept from the Euclidean set and which ones are appended.

We keep the patterns for the legs, and for the bases. In the patterns for the joining elements, we keep the vertex and mid-point patterns.

The new patterns are the information patterns. We do not have the neutral or passive elements: they are the patterns of the mantilla and the isoclines. We have new information elements which we display below, in figure 13. In the joining patterns, we have new corners which replace those of the Eucldean model. They are displayed by figure 14. Due to the new complexity of the horizontal signals, their names are a bit different from those of the Euclidean model. The names of these patterns now have the following syntax: $\boxed{H\gamma\xi\pi}$, with $\gamma$, $\xi$ and $\pi$ with their usual meaning. Of course, we keep the same name of the form $\boxed{H\gamma}$, with $\gamma \in \{g, y\}$, for the green and the yellow signals.

We have specific names for the scent. They have the form $\boxed{S\omega\nu\kappa}$ where $\omega \in [0..4]$, $\nu \in \{b, w\}$ and $\kappa \in \{s, c, f\}$. The index $\omega$ is the number of the isocline of the tile modulus 5. The letters $b$ and $w$ correspond to black and white, the status of the tile as a node in the Fibonacci carpet. The symbols $s$, $c$ and $f$ correspond to the start, the continuation or the final point of the signal.

The names are:

$Hblu$ $Hbru$ $Hrlu$ $Hrru$ $Hbll$ $Hbrl$ $Hrll$ $Hrrl$
$Hg$ $Hy$

and, for the horizontal red and blue signals and, for the scent they are as follows:

$S0bc$ $S0bs$ $Sibc$ $Siwc$ $S0bf$ $S0wf$

with $i \in [0..4]$. The corresponding tiles are displayed by figures 13, the patterns for the scent being excepted, as we have not enough room for that: see [10].

*c.* info patterns:

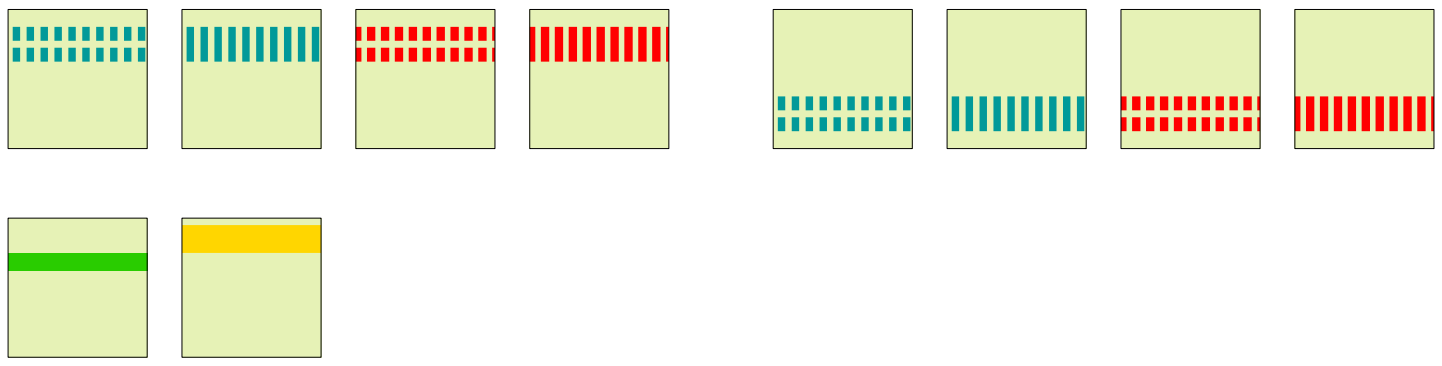

**Figure 13** *The elementary patterns for the horizontal signals: red, blue, yellow and green. Note the new elements.*

For the join patterns, the names are the same as in the Euclidean model but the shape of the pattern is different for the corner, as already mentioned, see figure 14 where we indicate the new patterns only.

the corners:

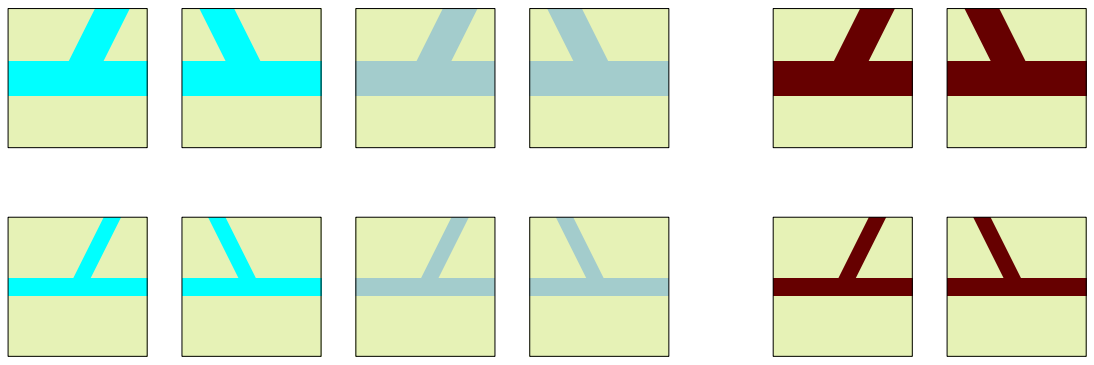

the joing tiles for horizontal signals:

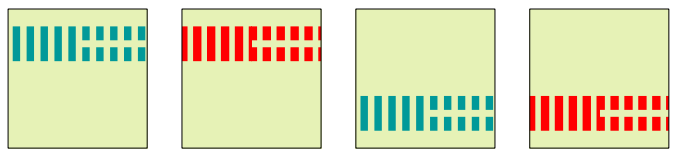

**Figure 14** *The elementary patterns for the join tiles.*

*For the crossing of legs and bases, we have all possible cases for blue and red legs, the blue-0 case being excluded by definition, and also all possible cases for the basis, this time the blue-0 case being taking into account.*

At last, we have the joining of horizontal signals which are also new signals: a right-hand side signal coming from the left meets a left-hand side signal coming from the right. Their joining is permitted and so, the corresponding tiles are provided. They are of the form $\boxed{J\gamma\pi}$, namely:

$$Jbu \qquad Jru \qquad Jbl \qquad Jrl$$

It is not difficult to see that $J\gamma\pi = \mu_L(H\gamma r\pi) + \mu_R(H\gamma l\pi)$.

*The tables of the tiles*

We define the tiles by grouping them as follows: vertices, first half of the legs, mid-distance points, second half of the legs, corners, bases, horizontal signals. To this purpose, we assemble the previous patterns. To simplify the expressions, we shall make use of the abbreviations which we have just introduced in the previous part of this sub-section.

We gather the results of the study into tables which will also allow us to count the prototiles, namely tables 3 and 4 for the description of the prototiles and table 5 for the counting.

We have no room here to enter a precise study of each group. We simply give the tables with the minimal amount of information to make them understandable.

The goal of the formulas is to represent all the tiles we need from the elementary patterns we have defined. They have to be considered as purely syntactic elements, especially when using them for counting. We refer the reader to [10] for the precise construction of the tables.

In many places, we shall use $\epsilon_i$'s as in the Euclidean case. We sjall define them as follows, in order to have an easier interpretation of the formulas.

The main lines of the construction allow to immediately understand the meaning of the following conditions:

$$\epsilon_y = \mathbf{1}_{\{\text{yellow row}\}},$$
$$\epsilon_c = \mathbf{1}_{\{\text{covered basis}\}},$$
$$\epsilon_s = \mathbf{1}_{\{\text{scent received}\}},$$
$$\epsilon_a = \mathbf{1}_{\{\text{accompanying horizontal signal}\}}.$$

They are used in formulas $(B_{b_0})$ and $(B_{b_n,r})$, and also in a few other formulas where the green or the yellow signal may occur.

We have the following result which is proved in [10].

**Lemma 20** *The set of prototiles defined by tables* 3, *and* 4 *force the construction of the tiling of the interwoven triangles in the hyperbolic plane.*

Now, our preparation is completed. But before turning to the description of the computing areas, let us look at a point which will again appear in the proof of the main theorem.

Indeed, the construction allows to realize all possible models of the abstract brackets. In particular, let us have a look at the butterfly model. In this case, there is a row of the mantilla, call it 0, which contains an infinite green signal. In fact, this row is on an isocline 15, and the line is the centre of the latitudes of infinitely many infinite towers of phantoms. Now, by the structure of the tiles, such a green line is never left alone.

| case of | formula |
|---|---|
| blue-0 bases<br><br>$(B_{b_0})$ | $Bb_0\tau + \epsilon_a.\Big(\mathbf{1}_{\{\tau=\varphi\}}.(a.Hb\xi_1 l + \overline{a}.Jbl)$<br>$+ \epsilon_c.(b.Hb\xi_2 u + \overline{b}.Jbu)$<br>$+ \epsilon_y.Hy + \overline{\epsilon_y}.(c.Hr\xi_3 u + \overline{c}.Jru) + \epsilon_s.S0\nu f\Big)$,<br>with $\epsilon_c \Rightarrow \overline{\epsilon_s}$, $a, b, c \in \{0, 1\}$ and the conditions $(a + b)(b + c)(c + a) > 0$, $\overline{\epsilon_c} \Rightarrow b$ and $\tau = t \Rightarrow c$.<br>denoted $B_{b_0}(\gamma, \tau, \xi_i, \epsilon_i, x)$ |
| simple blue and red bases<br><br>$(B_{b_n,r})$ | $B\gamma\tau + \mathbf{1}_{\{\tau=\varphi\}}.(a.H\gamma\xi_1 l + \overline{a}.J\gamma l)$<br>$+ \epsilon_c.(b.H\gamma\xi_2 u + \overline{b}.J\gamma u)$<br>$+ \epsilon_m.(Hg + \mathbf{1}_{\{\gamma=b_n\}}.Hy)$<br>$+ \overline{\epsilon_m}.(c.H\overline{\gamma}\xi_3 u + \overline{c}.J\overline{\gamma}u) + \epsilon_4.S0\nu f$,<br>with $\epsilon_s \Rightarrow \epsilon_m$ and $\epsilon_c \Rightarrow \overline{\epsilon_s}$, $a, b, c \in \{0, 1\}$, and the conditions $(a + b)(b + c)(c + a) > 0$, $\overline{\epsilon_c} \Rightarrow b$ and $\tau = t \Rightarrow c$.<br>denoted $B_{b_n,r}(\gamma, \tau, \xi_i, \epsilon_i, x)$ |
| vertices<br><br>$(V)$ | $V\gamma\tau + B\gamma\overline{\tau} + \mathbf{1}_{\{\tau=t\}}(\mu_L(H\gamma ll) + \mu_R(H\gamma rl))$<br>$+ \mathbf{1}_{\{\gamma\neq b_0\}}(Hg + \epsilon_s.S0bc + \mathbf{1}_{\{\gamma=b_n\}}.Hy)$<br>$+ \mathbf{1}_{\{\gamma=b_0\}}(\epsilon_y.Hy. + \overline{\epsilon_y}.H\overline{\gamma}\xi_1 u + \epsilon_s.S0bc$<br>$+ \overline{\epsilon_s}.(S0bs + \epsilon_c.H\gamma\xi_2 u))$ |
| mid-points<br><br>$(M)$ | $M\gamma\tau\xi + \mathbf{1}_{\{\tau=t\}}.(B\overline{\gamma}\tau_1 + \mu_{\overline{\xi}}(Hg) + \mu_\xi(H\gamma\xi u)$<br>$+ \mathbf{1}_{\{\tau_1=\varphi\}}.H\overline{\gamma}\xi_1 l + \mathbf{1}_{\{\gamma=r\}}.\mu_{\overline{\xi}}(Hy) + \epsilon_c.H\overline{\gamma}\xi u)$<br>$+ \mathbf{1}_{\{\tau=\varphi\}}.(B\gamma_1\tau_1 + Hg + \mathbf{1}_{\{\gamma_1=b_n\}}.Hy$<br>$+ \mathbf{1}_{\{\tau_1=\varphi\}}.H\gamma_1\xi_1 l$<br>$+ \epsilon_c.(\mathbf{1}_{\{\gamma_1\neq\gamma\}}.H\gamma_1\xi_2 u + \mathbf{1}_{\{\gamma_1=\gamma\}}.H\gamma\xi u))$<br>$+ \mathbf{1}_{\{\gamma=b_0\}}.\overline{\epsilon_c}.S0bf.$ |
| legs: first half, crossing phantom bases<br><br>$(L_u\varphi)$ | $L\gamma\tau u\xi + \mathbf{1}_{\{\gamma_1=\gamma\}}.$<br>$\Big(\mathbf{1}_{\{\gamma_1=b_0\}}.B_{b_0}(\tau_1, \xi_i, \epsilon_i)[\tau_1{:=}\varphi, \xi_2{:=}\xi])$<br>$+ \mathbf{1}_{\{\gamma_1\neq b_0\}}.B_{bn,r}(\gamma_1, \tau_1, \xi_i, \epsilon_i)[\tau_1{:=}\varphi, \xi_2{:=}\xi, \epsilon_m{:=}0])$<br>$+ \mathbf{1}_{\{\gamma_1\neq\gamma\}}.$<br>$\Big(\mathbf{1}_{\{\gamma_1=b_0\}}.(\mathbf{1}_{\{\tau=\varphi\}}.B_{b_0}(\tau_1, \xi_i, \epsilon_i)[\tau_1{:=}\varphi, \xi_3{:=}\xi]$<br>$+ \mathbf{1}_{\{\tau=t\}}.(\epsilon_y.(\mu_\xi(Hr\xi u) + \mu_{\overline{\xi}}.(Hy)$<br>$+ S0bf) + \overline{\epsilon_y}.(Hr\xi u + H\gamma_1\xi_1 u)$<br>$+ H\gamma_1\xi_2 l)$<br>$+ \mathbf{1}_{\{\gamma_1\neq b_0\}}.B_{b_n,r}(\gamma_1, \tau_1, \xi_i, \epsilon_i)[\tau_1{:=}\varphi, \xi_3{:=}\xi, \epsilon_m{:=}0])\Big)$ |

**Table 3** *Table of the formulas describing the prototiles, first part.*

| case of | formula |
|---|---|
| legs: first half, crossing triangle bases<br><br>$(L_u t)$ | $L\gamma\tau u\xi + \mathbf{1}_{\{\gamma_1=\gamma\}}.$<br>$\Big(\mathbf{1}_{\{\gamma_1=b_0\}}.B_{b_0}(\tau_1,\xi_i,\epsilon_i)[\tau_1{:=}t,\xi_2{:=}\xi,\epsilon_{c,s}{:=}1,0])$<br>$+\ \mathbf{1}_{\{\gamma_1\neq b_0\}}.B_{bn,r}(\gamma_1,\tau_1,\xi_i,\epsilon_i)[\tau_1{:=}t,\xi_2{:=}\xi,$<br>$\epsilon_{c,s,m}{:=}1,0,0])\Big)$<br>$+\ \mathbf{1}_{\{\gamma_1\neq\gamma\}}.$<br>$\Big(\mathbf{1}_{\{\gamma_1=b_0\}}.\Big(\mathbf{1}_{\{\tau=\varphi\}}.B_{b_0}(\tau_1,\xi_i,\epsilon_i)[\tau_1{:=}t,\xi_3{:=}\xi]$<br>$+\ \mathbf{1}_{\{\tau=t\}}.(B\gamma_1 t+\epsilon_c.H\gamma_1\xi u$<br>$+\ \epsilon_y.(\mu_\xi(H\gamma\xi u)+\mu_{\overline{\xi}}(Hy)$<br>$+\ \epsilon_s.S0bf)+\overline{\epsilon_y}.H\gamma\xi u)\Big)$<br>$+\ \mathbf{1}_{\{\gamma_1\neq b_0\}}.B_{b_n,r}(\gamma_1,\tau_1,\xi_i,\epsilon_i)[\tau_1{:=}t,\xi_3{:=}\xi,\epsilon_m{:=}0])\Big),$ |
| legs: second half<br><br>$(L_\ell)$ | $L\gamma\tau l\xi + \mathbf{1}_{\{\gamma_1=\gamma\}}.$<br>$\Big(\mathbf{1}_{\{\gamma_1=b_0\}}.B_{b_0}(\tau_1,\xi_i,\epsilon_i)[\xi_2{:=}\xi,\epsilon_{c,s}{:=}1,0])$<br>$+\ \mathbf{1}_{\{\gamma_1\neq b_0\}}.B_{bn,r}(\gamma_1,\tau_1,\xi_i,\epsilon_i)[\xi_2{:=}\xi,\epsilon_{c,s,m}{:=}1,0,0])\Big)$<br>$+\ \mathbf{1}_{\{\gamma_1\neq\gamma\}}.$<br>$\Big(\mathbf{1}_{\{\gamma_1=b_0\}}.(Bb_0\tau_1+\ Hb\xi_1 u+\mathbf{1}_{\{\tau_1=\varphi\}}.Hb\xi_2 l$<br>$+\ \mathbf{1}_{\{\tau=t\}}.$<br>$(\epsilon_y.(\mu_{\overline{\xi}}(Hy)+\mu_\xi(Hr\xi u))$<br>$+\ \overline{\epsilon_y}.Hr\xi u))$<br>$+\ \mathbf{1}_{\{\gamma_1\neq b_0\}}.B_{b_n,r}(\gamma_1,\tau_1,\xi_i)[\xi_3{:=}\xi,\epsilon_{c,s,m}{:=}1,0,0])\Big),$ |
| corners<br><br>$(C)$ | $C\gamma\tau\xi+\mathbf{1}_{\{\tau=\varphi\}}.H\gamma\xi_1 l+\mu_\xi(H\gamma\xi u)$<br>$+\ \mathbf{1}_{\{\gamma_1\neq b_0\}}.H\overline{\gamma}\xi_2 u$<br>$+\ \mathbf{1}_{\{\gamma_1=b_0\}}.\epsilon_a.(\epsilon_y.Hy+\overline{\epsilon_y}.H\overline{\gamma}\xi_2 u).$ |
| passive legs<br>$(L_{pz})$ | $L\gamma\tau\xi\pi$ |
| passive scent<br>$(S_{pz})$ | $Si\nu c$ |
| blank<br>$(W)$ | $W$ |

**Table 4** *Table of the formulas describing the prototiles, second part.*

| formula | colour | patterns | isoclines | mantilla | prototiles |
|---|---|---|---|---|---|
| $(B_{b_0})$ | blue-0 | 86 | phantoms: 0<br>triangles: 10 | 34 | 2534 |
| $(B_{b_n,r})$ | simple blue<br>red | 62<br>62 | 15<br>5, 15 | 34 | 6012 |
| $(V)$ | blue-0<br>simple blue<br>red | 24<br>4<br>4 | 0, 10<br>15<br>5, 15 | 1 | 36 |
| $(M)$ | blue-0<br>simple blue<br>red | 48<br>48<br>48 | 0, 10<br>15<br>5, 15 | 3 | 576 |
| $(L_u\varphi)$ | blue-0<br>simple blue<br>red | 154<br>64<br>648 | 0, 10<br>15<br>5, 15 | 3 | 1038 |
| $(L_ut)$ | blue-0<br>simple blue<br>red | 46<br>20<br>20 | 10<br>15<br>5, 15 | 3 | 318 |
| $(L_\ell)$ | blue-0<br>simple blue<br>red | 72<br>48<br>48 | 10<br>15<br>5, 15 | 3 | 648 |
| $(C)$ | blue-0<br>simple blue<br>red | 24<br>12<br>12 | 10<br>15<br>5, 15 | 3 | 180 |
| $(L_{pz})$ | any | 24 | 1-4, 6-9, 11-14, 16-19 | 3 | 1152 |
| $(S_{pz})$ | any | 2 | '1-4' | 34 | 144 |
| $(W)$ | any | 1 | 1-4, 6-9, 11-14, 16-19 | 34 | 544 |
| total | | 997 | | | 13,132 |

**Table 5** *The counting of the tiles.*

It is always accompanied by a basis. In this case, the basis never finds its legs and it never finds corners neither. If it is a basis of a phantom, it gives rise to infinitely many triangles which are infinite: this is clear from lemmas 16 and lemma 17, as legs created inside $T$ will be stopped by a triangle whose vertex appears inside $T$. If the basis is red, we have infinitely many infinite red triangles in which an infinite computation may take place in the case when the simulated Turing mahine does not halt.

Now, if the basis which accompanies the infinite green signal is a basis of a triangle, and of a red triangle, then it is possible, but not forced by the tiles, that there are infinite yellow signals on infinitely many rows: those which separate the latitudes of contiguous red triangles, see lemma 9. Again, we shall look again at this possiblity in the part devoted to the proof of the main theorem.

## 3.3 The new harps

At this moment, we have constructed what corresponds to the **skeleton** in the proofs of Berger and Robinson. Now, we are ready for the complete description of the computing areas.

The computing areas of our construction are the red triangles, whatever the generation. Accordingly, we forget the blue trilaterals, the red phantoms and the synchronization signals outside the basis of the red triangles.

Remember that in [9], we have considered the harps as implementations of a space time diagram of a Turing machine, as indicated in section 2.3.

In our new setting, the horizontal paths for the Turing head will be the free rows of the considered red triangle. What we now need is to describe the verticals. In order to not confuse them with the legs of the trilaterals which we also called verticals up to now, we shall call **perpendiculars** the new object which we now define which will play the rôle of a vertical in a space-time diagram: each one of our perpendiculars will contain the history of a given square of the Turing tape of the simulated machine.

For that purpose, we notice that the red triangles are set in trees of the mantilla whose borders do not meet. Now, the trees of the mantilla belong themselves to sectors of the mantilla and, from [9], we know that the borders of a sector do not meet the borders of a tree of the mantilla rooted in the centre of the head of the sector. We shall call **perpendicular** any ray which starts from the border of a tree and which follows the border of a sector. Accordingly, perpendicular never meet a leg of a trilateral.

Consider a red triangle $T$. We shall consider that the tiles which are at the intersection of its right-hand side leg and its left-hand side leg with a free row are the squares of the Turing tape which we consider as bi-infinite. Fix a square 0 on the Turing tape, which we may consider as the initial position of the head. We associate this square of the Turing tape to the vertex of $T$. We may view the right-hand side leg as the squares with a positive address and the left-hand side one as the squares with a negative address. Note also that we shall assume, without loss of generality, that the instructions of the Turing machine always indicate an actual move: either to right or to left.

Now, the evolution in time of the squares will be performed by the perpendicular which can be drawn from the place of the considered square on the border of the triangle. It takes the nearest border of a sector inside the triangle. We simply notice that, due to the period 3 of the tile pattern on both the left-hand side and the righ-hand side border of a tree of the mantilla, we may find an **8**-centre within two or three isoclines below the free row and, from this centre, the signal will follow the ray of **8**-centres, repeating the same period of three tiles. This is easy to perform: the pictures of figure 15 illustrate the implementation of the perpendiculars.

The new tiles needed for that are: tiles of the border of a tree, one or two intermediate tiles and then the tiles along a ray passing through **8**-centres. In [10] we give the exact description of the possible paths.

From the right-hand side part of figure 15, we can see that the concerned signal starts from the centre of a tile and goes to the mid-point of an edge 4 or 5. Later, the signal reaches an **8**-centre either through one or through two tiles. Once the **8**-centre is reached, a period is repeated which is indicated in the last three rows of the first column.

At last, inside $T$, a perpendicular never meets a leg of a triangle nor a basis of triangles which are inside $T$. The perpendicular will eventually meet the basis of $T$ and only it.

The computing signal runs along yellow rows and the just defined perpendiculars. It always conveys a state of the machine and it also indicates the direction of the motion of the head, using lateral signals, following the same patterns as those used for horizontal blue and red signals. The signal has the form $T\sigma\xi$ with $\xi = r$ when the signal moves to the right and $\xi = l$ when it moves to the left. The meaning of $\sigma$ is defined a bit later.

When it meets a perpendicular, which always indicates the current content of a square of the Turing tape a new instruction is performed and we have two cases. Either the move of the head is left unchanged by the instruction, or it is changed. In the first case, the signal goes on on the row without changing its laterality, unless the meeting with the perpendicular occurs at a tile which is on the border of a tree. In this latter case, the signal behaves as if the new move takes the opposite direction with respect to the former move: it goes down along the perpendicular. After the meeting, if the signal has to go down, it indicates the new content of the square, the new state of the head and the new direction it will have to take. When the signal reaches the next yellow row, the content continues to go down along the perpendicular and the state goes on its way this time along the row, in the direction indicated by its own signalization. In this way, the next perpendicular which will be met will be the expected one: it will hold the content of the concerned square of the Turing tape.

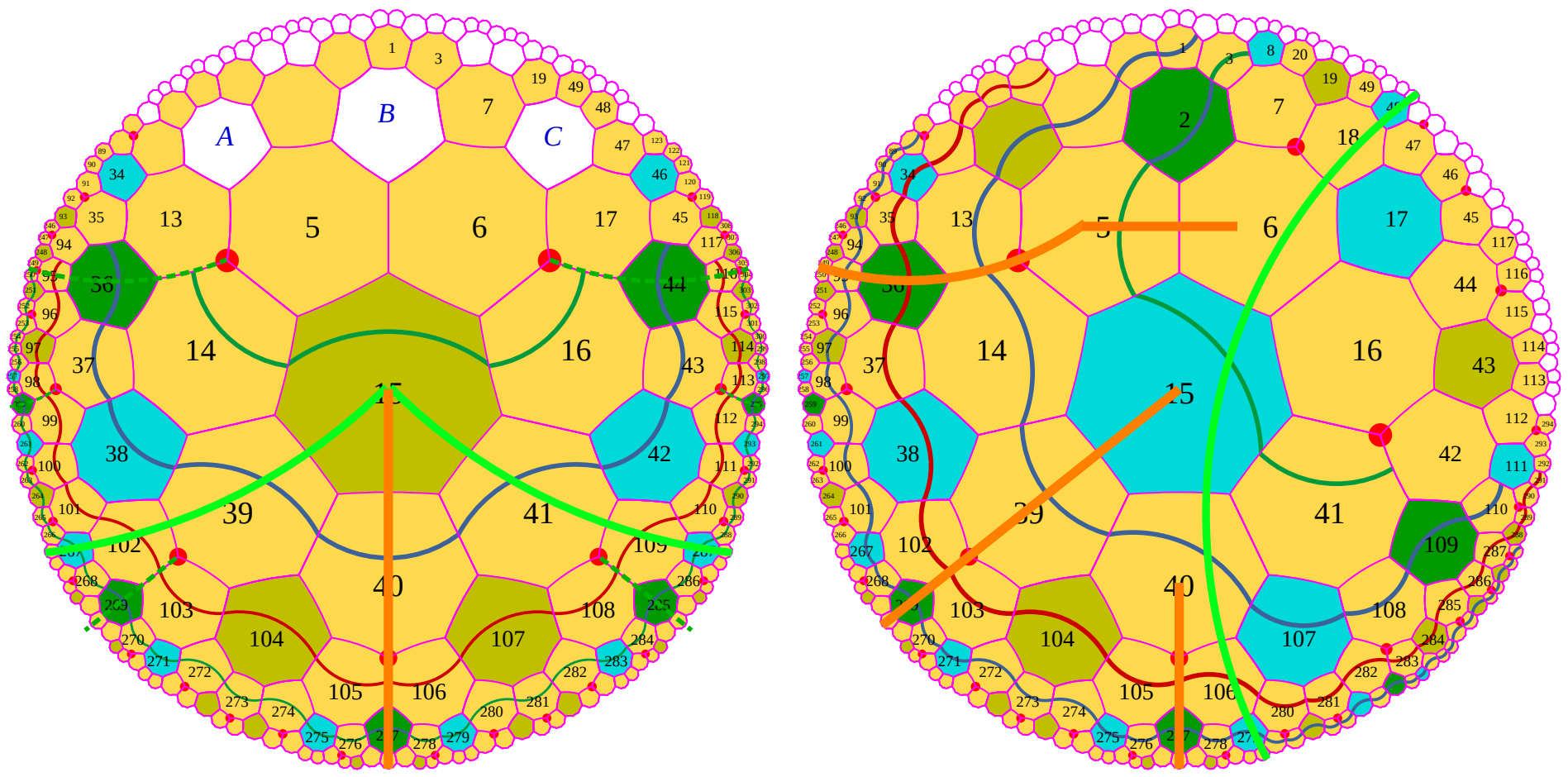

**Figure 15** *The perpendicular starting from a point of the border of a triangle which represents a square of the Turing tape.*

*On the left-hand side: the case of the vertex. On the right-hand side, the three other cases for the right-hand side border are displayed on the same figure.*

Note that our general scheme differs a bit from that of [9]. With respect to that scheme, the present introduces a delay in the transfer of the new state. In case of a change in the direction of the motion of the head, the signal first follows the perpendicular, accompanying the new content of the square and then, when it finds the next yellow row, it goes on the row in the direction indicated by its signal. This is why we provide new tiles for the computation signals which are displayed by figure 17 and their use in accordance with the previous construction signals is precisely described by table 6. The figure indicates the signals in the square format for the convenience of the reader. Below, in figure 16, we first give the names of the new signals: they are of the form $T\sigma[\xi]$, as already mentioned. Here $\xi \in \{l, r\}$ indicates the laterality and $\sigma \in \{i, b, s, t, m, e, c, u, hu, hc\}$ indicates the nature of the operation performed by the tile. Thus, $i$ indicates the start of the signal which only matches with the vertex of a red triangle. Next, $b$ indicates a border tile: $Tbi$ stands for an initialization of the corresponding Turing square, $Tbl\xi$ indicates the two possible cases when the computing signal arrives at the left-hand side leg, while $Tbr\xi$ does the same for the right-hand side leg. These tiles execute an instruction. Then, $s$ and $t$ simply convey the information: $s$ for the state of the head of the Turing machine when it is conveyed by a yellow row, $t$ for the content of the square when it is conveyed by a perpendicular. The letters $m$ and $e$ refer to performing an instruction: $m$ indicates the moment when a state signal meets a square cell. In fact, a yellow row crosses a perpendicular. At the intersection, we have a tile $Tm\xi$, depending on the motion of the head. The tile executes the instruction corresponding to the symbol conveyed by the perpendicular and the state conveyed by the row. As output, the tile indicates the new content of the tape, the new state of the head and the direction of the next motion of the head.

There are two cases, depending on whether the new direction is different or not from the current one. If it is not different, this is indicated by the tiles $Tu\xi$ which also perform the instruction. If it is different, the tiles $Tm\xi$ execute the instruction. Both the new state, the new content and the new direction are conveyed by the perpendicular through the tiles $Tc\xi$. When the perpendicular meets the next yellow row, the state and the content are separated by the tiles $Te\xi$. Afterwards, the tiles $Ts\xi$ and $Tt$ convey the new information.

| | | | | | | | |
|---|---|---|---|---|---|---|---|
| $Ti$ | $Tsl$ | $Tsr$ | $Tt$ | $Tbi$ | | | |
| $Tbll$ | $Tblr$ | $Tbrl$ | $Tbrr$ | $Tml$ | $Tmr$ | $Tul$ | $Tur$ |
| $Tel$ | $Ter$ | $Tcl$ | $Tcr$ | | | | |
| $Tdl$ | $Tdr$ | $Thul$ | $Thur$ | $Thcl$ | $Thcr$ | | |

**Figure 16** *The names of the computing meta-tiles.*

At last, but not least, the tiles $Td\xi$, $Thu\xi$ and $Thc\xi$ manage the interruption and the halting case.

The interruption case is defined by the tiles $Td\xi$ which are used when the computing signal, being on a perpendicular, meets the basis of the red triangle where the computation is performed. As the corresponding signal may convey a left-hand side or a right-hand side new direction, we have two tiles: $Tdl$ and $Tdr$.

The halting case is managed by the tiles $Thu\xi$ and $Thc\xi$. The halting state appears in a tile which performs an instruction: $Tm\xi$ or $Tu\xi$, depending on whether the new direction of the move of the machine head is different or not from that of the current move. In both cases, we consider that the halting state is black. Now, after a tile $Tm\xi$, a tile $Tc\xi$ comes and, after a tile $Tu\xi$, it is a tile $Ts\xi$. We decide that no tile of this kind bears a black sign for the state. The only tiles with a black sign for the states are the tiles $Thu\xi$ which match with the tiles $Tu\xi$ and the tiles $Thc\xi$ which match with the tiles $Tm\xi$ only. Now, no other tiles may match with the tiles $Thu\xi$ and $Thc\xi$ which, accordingly cause the tiling to stop. Note that for the tiles $Thc\xi$ and $Thu\xi$ it is important that the tiles can be neither rotated nor reflected.

Figure 17 illustrates these tiles, giving them in the order given above.

Now, the implementation of the harp is completed.

We have just to count how many tiles are appended by those of figure 17. Note that these tiles are different from those of the skeleton. Indeed, the tiles of the figure are not an abbreviated representation of prototiles, but an abbreviate representation of *variables* for prototiles, let us call them **meta-tiles**. And so, they should be counted separately.

To facilitate the couting, we introduce a table in the same way as the ones we introduced to list the tiles needed by the skeleton, see table 6.

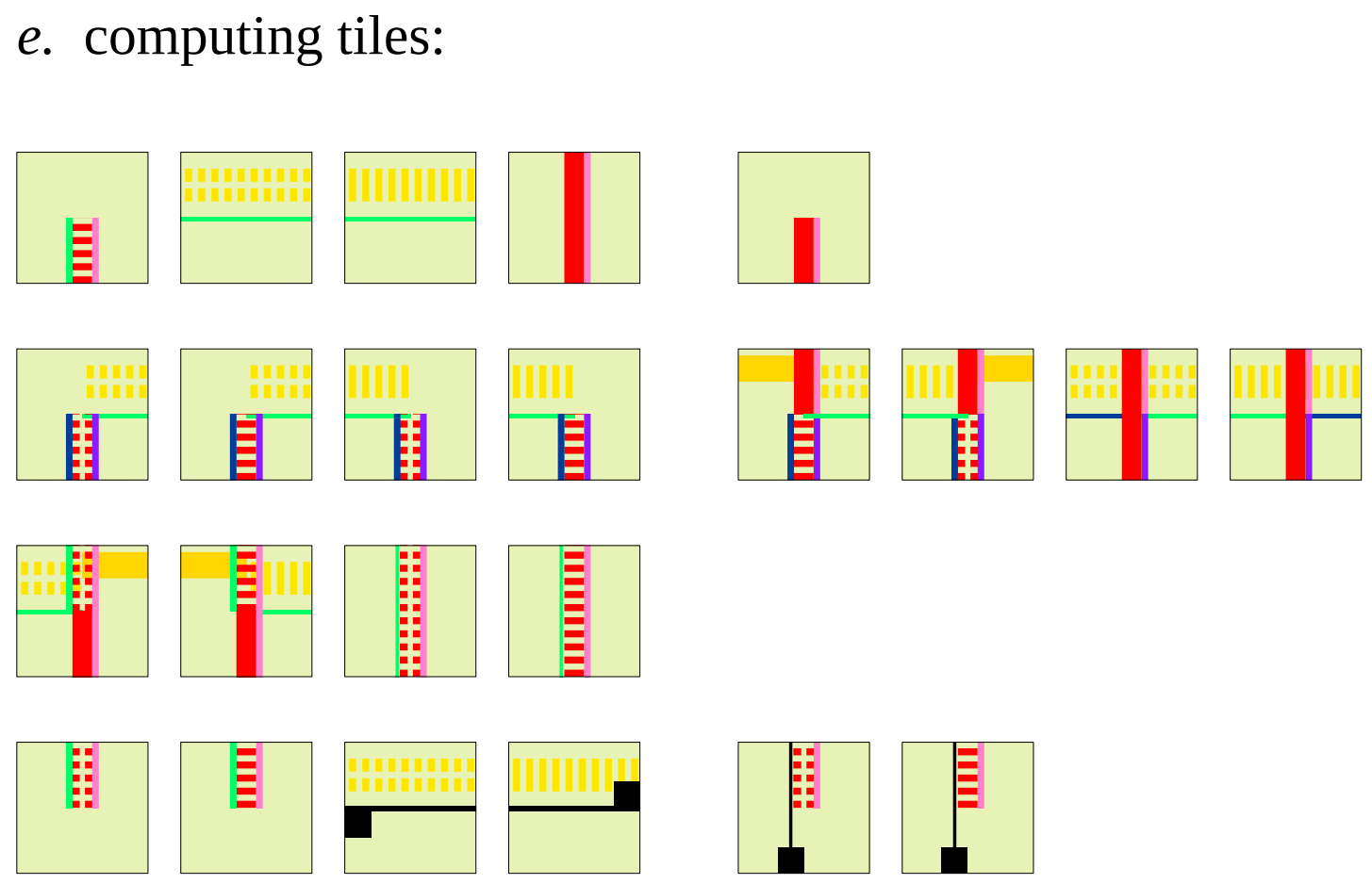

**Figure 17** *The computing tiles.*

The details of the couting are given in [10]. It is based on the analysis of tables 3 and 4 from which we can extract the number of patterns which bear yellow signal and on the analysis, in [10] of the number of tiles used for the starting tiles of a erpendicular.

Summing up all the cases which appear in table 6, we find out 13540 meta-tiles.

| type | formula | patterns | # |
|---|---|---|---|
| $B_{yellow}$ tiles : | | | |
| 'pure' basis | $a.Ts\xi + \overline{a}.Thu\xi + b.(B_{b_0}) + \overline{b}.(B_{b_n})$ | 23+23 | 6256 |
| corner | $a.Ts\xi + \overline{a}.Thu\xi + (C)$ | 12 | 72 |
| mid-point | $a.Ts\xi + \overline{a}.Thu\xi + (M)$ | 36 | 432 |
| legs: | | | |
| first half | $a.Ts\xi + \overline{a}.Thu\xi + b.(L_u\varphi) + \overline{b}.(L_u t)$ | 238+60 | 3576 |
| second half | $a.Ts\xi + \overline{a}.Thu\xi + (L_\ell)$ | 100 | 1200 |
| starting point of perpendiculars: | | | |
| inside the yellow row: | | | |
| vertex | $Ti + a.Ts\xi + b.Thu\xi + (V)$, $a + b \leq 1$ | 12 | 120 |
| ends of yellow rows: | | | |
| first half | $a.Tbi + \overline{a}.Tb\xi\xi_1 + b.(L_u\varphi) + \overline{b}.(L_u t)$ | $2 + 8$ | 90 |
| second half | $a.Tbi + \overline{a}.Tb\xi\xi_1 + (L_\ell)$ | 2 | 18 |
| mid-point | $a.Tbi + \overline{a}.Tb\xi\xi_1 + (M)$ | 12 | 216 |
| the tape content | $a.Ti + b.Tt + c.Tc\xi + d.Thc\xi + (P)$, $a + b + c + d = 1$ | 90 | 720 |
| execution | $a.Tm\xi + \overline{a}.Tu\xi + \{B_{yellow} \cap (P)\}$ | 46 | 552 |
| splitting the Turing output | $Te\xi + \{B_{yellow} \cap (P)\}$ | 46 | 276 |
| stopping by meeting the basis | $Td\xi + Brt$ | 6 | 12 |

**Table 6** *The precise description of the computing tiles and their counting.*

The counting is summed up in table 6, above.

## 3.4 The proof

The proof follows the same pattern as Berger's and Robinson's proofs. In the case when the simulated Turing machine does not halt, whatever the size of the computing areas, these region can be tiled.

In the case when the simulated Turing machine halts, we have to consider the possibility of starting the construction from any tile.

As in Berger's and Robinson's case, we can first consider the tiles omitting the computing signs as well as those which eventually concern the interwoven triangles. And so, we first know in which part of the mantilla we are. Moreover, the isocline number tells us also where we are. This means that within a ball of radius at most 20, we find an active seed of the generation 0. We shall find also what is needed to start the generation 0 and the generation 1 on any length. Accordingly, from lemma 20 and the algorithm 4.2.2, the construction of the interwoven triangles is step by step forced. Now, from the properties of the interwoven triangles, as all of them occur at an appropiate latitude at least once, this provides computing areas with a size, as long as desired. Consequently, we shall find an area where the construction will be stucked by the halting of the Turing machine.

It remains to come back to the case of the butterfly model for which we have noted a few particularities after the proof of lemma 20. The case when a red basis occurs on the line 0 of the butterfly model is of interest for us. We noted that this case may happen and that it splits into two sub-case: we have a basis of a phantom or a basis of a triangle.

In the first case, as already noted, we have infinitely many infinite red triangles yelding infinite computations. If the simulated Turing machine halts, the construction of such triangles will be stopped at some point, possibly before other triangles, as they are bigger. If the simulated Turing machine does not halt, then the construction of such triangles go on endlessly and it does not interfer with the finite triangles which they contain. And so, in this case, the tiling can be completed.

In the second case, when the basis is that of a triangle, we have infinitely many phantoms below, which bring in nothing from the point of view of the computation. But, as noted, the presence of the basis, may involve, above, infinitely many infinite yellow rows, running across the hyperbolic plane, from left to right on this row. As these rows do not meet legs of the red triangle associated to this infinite basis, if they contain states, we may think that they are not connected between them. However, me may place verticals, as what is needed is a ray of **8**-centres and so, a computation may takes place with no starting point.

First, note that such a phenomenon happens above the infinite basis. Accordingly, correct computations occur below in the finite triangles and so, if the halting occurs, it will be detected.

Now, if the simulated machine does not halt, as we just need to construct one solution, we may proceed to a few choices. We may rule out the butterfly model: this means that each time we choose the row of the triangles of a new generation, we perform the choice accordingly to the previous ones in order, for instance, to guarantee that each row of the mantilla corresponds to a point which is infinitely many times covered by the active intervals.

But we may also decide to constuct the butterfly model only, in the case when the basis is that of a phantom. In this case this single construction provides us with a correct test: if the simulated machine does not halt, this provides us with a construction. If the simulated machine halts, the construction will be stucked and, from what we have seen before, we know that, starting from any tile, we shall find a big enough triangle in which the computation will halt, blocking the tiling.

And so, the main theorem is proved.

As a corollary, we can improve the construction of [9] to the following result:

**Corollary 2** *There is a finite set $\mathcal{S}$ of tiles such that there is a non-recursive way to tile the hyperbolic plane with copies of $\mathcal{S}$ but no recursive way.*

# Conclusion

I would like to conclude on two aspects of the construction.

The description of section 2, mainly the algorithm of section 2.3 and its preparation are in the style of signal drawn on the mantilla.

We have seen, in section 3.2.2, that this can be replaced by an algorithm in the style of [9] but we had to resign to the principle according to which once a tile is placed, it cannot be removed. We replaced it by a weaker version: it may be changed, but at most finitely many times. Indeed, we noticed that from time to time, later generations introduce something new in already settled domains without basically changing them. Note that this does not contradict the definition of the existence of at least one solution. Also note that the same phenomenon occurs with Robinson's construction in his proof.

I would like to indicate several things.

The first is that this solution is not unique, for sure. Of course we have continuously many realizations of the tiling: twice in some sense as there are continosly many realizations of the mantilla and continuously many models of the abstract brackets. I do not mean that. In fact there are possible variants of some points of the construction. As an example, it is possible that the yellow signal is useless. Perhaps horizontal blue signals not mixed with red ones are enough for our purpose. However, the yellow signal makes things easier and does not bring in too many tiles. Also, it simplifies things for the computation part of the proof. Another point is the use of the scent. Another variant could consist in using the perpendiculars as defined in section 3.3. We could restrict ourselves with the meeting of the perpendicular issued from the root of the tree with the mid-distance row to define the green signal. In this case, we have to draw perpendicular also from the vertex of a phantom. And then we have to define again a selection principle. The solution of the scent seemed to us a bit simpler. Now, a more important question can be raised: after all, why to restrict the number of active seeds? Why not to decide that all seeds are active? In the present setting of the proof, the advantage of a rather strong restriction of the active seeds is that we better see what happens. In particular, the situation raised by missing trilaterals is more clearly handled. But, in fact, it is perhaps not needed to have strong restrictions. The basic principle of our construction could be stated as: *something exists because the objects of the previous generations which it contains are necessarily present.* And so, the restrictions can be relaxed as long as this principle is remains valid. Perhaps it is possible to remove any restriction.

The basic tiling can also be questioned: why the mantilla and not simply the underlying $\{7, 3\}$ tiling or, the pentagrid, for instance?

The point is that in the mantilla, we have a way to define the computing verticals which is really very efficient: we know that the verticals which we defined, the **perpendiculars** defined in section 3.4, will never meet the legs of a triangle. Accordinlgy, dropping a vertical from any tile of the border of a tree $T$ guarantees that the vertical will never meet a tree which is rooted inside $T$. It can be argued that the mantilla is not the single tiling where the isocline property occurs. In a simple Fibonacci carpet, this basic property also occurs. But there, things are not that simple: what to choose in order to construct the triangles and the verticals? We need Fibonacci trees. The way we defined them in the mantilla can of course be adapted here. Now, are we sure that the trees will keep the property that their legs do not intersect, as observed in the mantilla? This is not easy to prove and it cannot be at first glance. It needs further investigations. And now, how to define verticals? The suggested investigations have also to take this aspect in consideration.

It is plain that the huge number of tiles is due to the 20 degrees in the isoclines. But this is needed to fix the generations 0 and 1 because of lemma 3. Five is indeed the smallest distance from the root at which we can find other seeds. And so, to significantly reduce the number, it should be needed to change the basic tiling.

A last but not least point is what does the proof learn to us?

It seems to me that the main lesson of the proof is that the realm of absolute geometry is probably under-estimated. What shows the proof is that a construction which seems to be typically Euclidean is indeed not: it is purely combinatorial and it makes a very little use of the geometry, at least we do not have to bother about parallels.

And it is interesting to notice that Robinson's question about the importance or not of non-periodic constructions in the case of the hyperbolic plane can be answered positively. In this proof, the tiling is not periodic: no shift can keep the tiling invariant.

### Acknowledgment

I would like very much to thank both Chaim Goodman-Strauss and Jarkko Kari again for pointing at my error in my previous attempt to prove the theorem. I would also like to thank Leonid Levin for indicating me his paper I quoted about the abstract brackets. I wish also to thank very much three colleagues, André Barbé, Serge Grigorieff and Tero Harju, and again Chaim Goodman-Strauss for their moral support.